\documentclass[aps,prd,twocolumn,showpacs,superscriptaddress,nofootinbib]{revtex4}  
\usepackage{graphicx}  
\usepackage{dcolumn}   
\usepackage{bm}        
\usepackage{amssymb}   
\usepackage{setspace}
\usepackage{hyperref}
\usepackage{textcomp}
\usepackage{amsmath}
\usepackage{amsfonts}
\begin{document}

\title{Semiclassical and quantum behavior of the Mixmaster model in the \textit{polymer} approach}

\author{Orchidea Maria Lecian}
\email{lecian@icra.it}
\affiliation{Dipartimento di Fisica (VEF), P.le A. Moro 5 (00185) Roma, Italy}

\author{Giovanni Montani}
\email{giovanni.montani@frascati.enea.it}
\affiliation{Dipartimento di Fisica (VEF), P.le A. Moro 5 (00185) Roma, Italy}
\affiliation{ENEA-UTFUS-MAG, C.R. Frascati (Rome, Italy)}

\author{Riccardo Moriconi}
\email{moriconi.phys@gmail.com}
\affiliation{Dipartimento di Fisica (VEF), P.le A. Moro 5 (00185) Roma, Italy}

\date{\today}

\begin{abstract}
We analyze the quantum dynamics
of the Bianchi Type IX model, as described
in the so-called polymer representation of quantum mechanics,
to characterize the modifications that
a discrete nature in the anisotropy variables
of the Universe induces on the morphology
of the cosmological singularity.
We first perform a semiclassical analysis,
to be regarded as the zero${}^{th}$-order approximation
of a WKB (Wentzel-Kramers-Brillouin) approximation of the quantum dynamics,
and demonstrate how the features of polymer quantum mechanics are
able to remove the chaotic properties of
the Bianchi IX dynamics.
The resulting evolution towards the cosmological singularity
overlaps the one induced, in a standard Einsteinian
dynamics, by the presence of a free scalar field.
Then, we address the study of the full quantum dynamics of this model
in the polymer representation and analyze the two cases, in which the
Bianchi IX
spatial curvature does not affect the wave-packet
behavior, as well as the instance, for which it plays the role of
an infinite potential confining the dynamics
of the anisotropic variables.
The main development of this analysis consists
of investigating how, differently from the standard
canonical quantum evolution,  
the
high quantum
number states are not preserved arbitrarily
close to the cosmological singularity.
This property emerges as a consequence,
on one hand, of the no longer chaotic features of the
classical dynamics (on which the Misner analysis
is grounded), and, on the other hand, of
the impossibility to remove the quantum
effect due to the spatial curvature.
In the polymer picture, the quantum evolution
of the Bianchi IX model remains always significantly
far from the semiclassical behavior, as far as both the wave-packet spread and
 the occupation quantum numbers are concerned. As a
result, from a quantum point of view, the Mixmaster dynamics
loses any predictivity characterization for
the discrete nature of the Universe anisotropy.
\end{abstract}

\maketitle

\section*{INTRODUCTION}
\label{intro}
Even though the thermal history of the Universe is
properly described by the homogeneous and isotropic
Robertson-Walker cosmological model \cite{kolbeturner}, up to the very
early stage of its evolution, reliably since
the inflationary phase has been completed, the
nature of the cosmological singularity is
a general property of the Einsteinian dynamics,
suggesting the necessity of relaxing the high
symmetry of the geometry that characterizes the
Big-Bang. A very valuable insight on dynamical
behaviors more general than the simple isotropic case, especially
in view of the classical features of
the initial singularity and of its quantum ones, is offered by the Bianchi
type IX model\cite{bklref0}-\cite{bklref8}, essentially for the following three
reasons: i) this model is homogeneous, but
its dynamics possesses the same degree of generality as
any generic inhomogeneous model ii) the canonical
quantization of this model can be performed via
a minisuperspace approach, which reduces the
asymptotic evolution near the singularity to
the well-defined paradigm of the 'particle in a box'
iii) the late (classical) evolution of the model is
naturally reconcilied, with or without the inflation\cite{montani-kirillov}\cite{montani-kirillov2},
to that of a closed isotropic Universe\cite{russi}.

Two years after the derivation of the Wheeler-DeWitt
equation, C.W.Misner applied this canonical quantization approach to the Bianchi IX model,
which, in the Hamiltonian version he had just provided (restating the oscillatory regime derived
by Belinski-Khalatnikov-Lifshitz),
constitutes a proper application.

The main result of the Misner quantum analysis of
the Bianchi IX dynamics is to provide a brilliant
demonstration of the phenomenon for which very high occupation numbers are
preserved in the evolution towards the cosmological singularity.
This result is achieved by using the chaotic
properties of the classical Bianchi IX dynamics
(the Mixmaster model) and by approximating the
potential well, in which the Universe-particle
moves, by a simple square box, instead of
the equilateral triangle it indeed is.

Many subsequent studies have been pursued on
such a quantum dynamics, both in the Misner
variables \cite{misnermixmaster}, as well as in other frameworks
(as the Misner-Chitr\'e scheme\cite{misnercitre},\cite{misnercitre2}), but the
original semiclassical nature of the cosmological
singularity, when considered in terms of high occupation
numbers, still remains the most striking prediction of
the canonical quantum dynamics provided by the
Mixmaster model.

In the Misner variables, the dynamical problem
is reduced to a two-dimensional scheme, in which
the role of the time variable is played by the Universe
volume, while the two physical degrees of freedom
are represented via the Universe anisotropies.
Such a picture is elucidated by an ADM (Arnowitt-Deser-Misner \cite{ADM}) reduction of the variational principle
(based on the solution of the superHamiltonian constraint),
but its physical significance emerges in a very
transparent way, already in the direct Hamiltonian
approach to the dynamics.

Here, we address the quantum analysis of the Bianchi
IX model by using the superHamiltonian constraint,
which, via the Dirac prescription, leads to the
Wheleer-DeWitt equation; this procedure is
expectedly fully consistent with the Schr\"odinger-like
dynamics following from the ADM reduction of the Dirac
constraint.
The new feature we introduce by the present study is
the discrete nature of the anisotropic degrees
of freedom, by a two-dimensional
quantization approach, based on the so-called \textit{polymer representation of quantum mechanics}. The use of such a modified quantization
scheme is justified by the request that a cut-off in the spatial scale,
as expected at the Planckian level, would induce a corresponding
discrete morphology in the configuration space. The
reason for retaining the isotropic Misner variables,
connected to the Universe volume, as a continuous ones,
relies on the role time plays in the dynamical scheme.


This way, we discuss first the semiclassical behavior of the model,
to be regarded to as the zero${}^{th}$-order approximation of a WKB expansion of
the full quantum theory.
By other words, we analyze the classical behavior of
a modified Hamiltonian dynamics of the Mixmaster model,
based on the prescriptions fixed by the classical limit
of the new paradigm. In this respect, we remark that the
typical scale of the polymer discretization is not directly related
to the value of $\hbar$; the classical limit for this
quantity approaching zero still constitutes a modification of
the Einsteinian classical dynamics. Such a semiclassical study is
relevant to the interpretation of the full quantum behavior of
the system, especially for the analysis of localized wave packets.
The main result of this semiclassical analysis is the
demonstration that the chaotic structure of the asymptotic
evolution of the Bianchi IX model to the cosmological
singularity is naturally removed by a dynamical mechanism
very similar to the one induced on the same dynamics by the introduction of
a free massless scalar field.
In the limit of small values of the polymer lattice
parameter, we calculate the modified reflection law
for the point-Universe against the potential walls, which are due to the
spatial curvature.
For the general case, we provide a precise description
of how the bounce against these walls is avoided and
of the condition that the free-motion parameters of the model
must satisfy for it to take place.

The absence of chaos in the semiclassical behavior of
the polymer Mixmaster model prevents us from directly implementing the Misner procedure, which is basic to its description
of the states approaching the singularity with very high occupation
numbers. We are therefore lead to analyze separately two cases:
one in which the potential walls can be neglected in the quantum
evolution, such that we deal with free-particle wave packets, and one in which
when the potential walls play the role of a 'box', in which the point-Universe
is confined. In both cases, the behavior
of localized wave packets is analyzed to better understand the limit up to which the Misner semiclassical feature survives in
this modified approach. The main outcome we develop here is
to identify how the presence of the walls is, sooner or later, relevant
for the wave packets evolution, such that also the states
with high occupation numbers are accordingly obliged to spread close enough to the
singularity, simply because the potential box
destroys their semiclassical nature.
The direct comparison with the Misner result is
not possible because of the semiclassical behavior of the model, but the conclusion
of our analysis is otherwise very clear, because
there is no chance to built up a localized state that can
reach the singularity without bouncing against the potential
walls. This is because the conditions to fix a direction in
the configuration space, which would allow for a free motion, is indeed
time dependent and is, sooner or later, violated during the
evolution of the wave packets.
In this scheme, there is no possibility to retain
 the semiclassical features in the quantum description, and
 we can therefore claim that the singularity of the Mixmaster
Universe, as viewed in the present polymer representation, can
not be described by semiclassical notion, as for
the Einsteinian oscillatory regime of the expanding
and contracting independent directions, and that even its
quantum relic, i.e. the occurrence of
 high occupation numbers close to
the singular point, is removed in a discrete quantum
picture for the anisotropy degrees of freedom.

The relevance of this result is enforced by observing that, as
 investigated in \cite{montani-kirillov}, an oscillatory regime cannot exist
before a real classical limit of the Universe is reached.
As a consequence, since a simple model for the cut-off physics
is able to cancel also the memory of semiclassical features
in the Planck era (as Misner argued in \cite{misner}), we are
lead to believe that the classical Mixmaster dynamics 
is not fully compatible with the quantum origin of the Universe, and
it is indeed a classical dynamical regime reached by the
system when the quantum effects are very small and the
physical and configuration spaces appear as bounded by continuous domains.

A certain specific interest for the implementation of
a polymer approach to the quantum dynamics of the Universe
was rised by the analogy of this quantum prescription to
the main issues of Loop Quantum Cosmology, which, in the
Minisuperspace, essentially reduces to a polymer treatment of the
Ashtekar-Barbero-Immirzi
variables, as adapted to the cosmological setting\cite{loop1}-\cite{loop3}.

The first Loop Quantum cosmological analysis
of the Bianchi IX model was provided in \cite{boj1},
where it was argued the non-chaotic nature
of the semiclassical dynamics. The main reason
of such non-chaotic behavior of the Bianchi IX
model must be determined in the discrete nature
of the Universe volume and, in particular in its
minimal (cut-off) value. In fact, asymptotically
to the singularity, the potential walls can no longer
arbitrarily growth and the point-Universe confinement
is removed. This analysis was based on the so-called
inverse volume corrections and properly accounts
for the induced semiclassical implications of
the Loop Quantum Gravity theory. Nonetheless, the
considered approach is based on a regularization scheme 
(the so called $\mu_0$ one) that is
under revision, in order to provide a consistent
reformulation of the Bianchi IX dynamics as done
in \cite{APS} for the the isotropic Robertson-Walker
geometry (the $\bar\mu$ regularization scheme).

A step in this direction has be pursued in \cite{wilson},
where the Loop quantum dynamics is rigorously restated by adopting 
the $\bar\mu$ regularization scheme, 
demostrating that, under certain circumstances, the
chaotic features of the Bianchi IX model are
removed again. However, this result holds only when
the quantum picture includes a massless scalar field,
able to remove the chaoticity even on a classical
Einsteinian level. The reason of this striking
difference in the two result obtained in these
two approaches, must be individualized in the
kind of semiclassical corrections discussed in \cite{wilson}.
Indeed, in this analysis only the so-called
Holonomy correction contributions are considered and they are unable
to induce on the dynamics the basic feature of
a volume cut-off scale. It is just this different
type of quantum corrections adopted to construct the
semiclassical limit, the reliable source of the
non-generic nature of the chaoticity removal.
Althought it is expected that inverse volume corrections can remove the 
chaotic behavior of the Bianchi IX model as in \cite{boj1}, nevertheless
this has not been explicitly demonstrated and it stands as a mere conjecture.

It is worth noting that a critical revision of
the Loop Quantum Cosmology picture of the primordial
Universe space, was presented in \cite{cian-mon},
where the necessity of a gauge fixing in implementing
the homogeneity constraint is required.
A consistent quantum reformulation of the dynamics of
an homogeneous model was then constructed in \cite{cian-ale},
which allows a semiclassical limit of the theory,
in close analogy to the full Loop Quantum Gravity theory.

Despite the polymer formulation of the
canonical approach to the minisuperspace mimics
very well some features of the Loop Quantum cosmology
methodology (de facto a polymer treatment of
the restricted Ashtekar-Barbero-Immirzi variables
for the homogeneity constraint), however, there is
a crucial difference between our result and such
recent Loop-like approaches.
In fact, we apply the polymer procedure to the
anisotropic variables only (the real degrees of
freedom of the cosmological gravitational field),
leaving the Universe volume at all unaffected
by the cut-off physics, in view of its time-like
behavior. The removal of the chaos, discussed here, is
therefore not related to the volume discretization
and it is also difficult to characterize its relation
with the Holonomy correction approach
(studying properties of the edge morphology more than
of the nodes, but non-directly reducible to the
anysotropy concept). In this respect, the present
result must be regarded as essentially an independent
one with respect to the ones actually available in
Loop quantum Cosmology.
\\
This paper is organized as follows.\\
In Section \ref{PQM}, we introduce the polymer representation of
quantum mechanics, by a kynematical and a dynamical point of view.
Then we analyze the continuum limit and conclude the Section illustrating
two fundamental examples of one-dimensional systems: the polymer free
particle and the particle in a box.\\
In Section \ref{MIX}, we review the principal (classical and
quantum) features of the Mixmaster model, as studied by Misner
in\cite{misner}.\\
Section \ref{SEMI} is dedicated to the study of the
polymer Mixmaster model, from a semi-classical point of view. In
particular, we analyze the modified relational motion between the
Universe-particle and the walls, and we derive a modified reflection
law for one single bounce against the wall.\\
In Section \ref{QUANTUM}, we build up the wave packets for the case when the wave
function of the Universe is related to a free polymer
particle and to a polymer particle in a square box, respectively.\\
Finally, Section \ref{NUM} is devoted to the numerical
integrations of the polymer wave packets and to the analysis on the
quantum numbers related to the anisotropy.\\
Concluding remarks complete the paper.
\section{The POLYMER REPRESENTATION OF QUANTUM MECHANICS}
\label{PQM}
To apply the modified polymer approach to the Mixmaster quantum dynamics, we briefly summarize the fundamental features of this modified quantization scheme. In particular, after giving a general picture of the model, we consider the two specific cases of the free particle and of the particle in a box, which are relevant for the subseguent cosmological study.
\subsection{Kynematical properties}
\label{Kin}
The Polymer representation of quantum mechanics is a non-equivalent representation of the usual Schr\"odinger quantum mechanics, based on a different kind of Canonical Commutation Rules (CCR). It is a really useful tool to investigate the consequences of the hypothesis for which the phase space variables are discretized.
\newline
For the definition of the kinematics of a simple one-dimensional system\cite{corichi}, one introduces a discrete set of kets $ |\mu_{i} \rangle$, with $\mu_{i} \in  \mathbb{R}$ and $i =1,...,N$. These vectors $ |\mu_{i} \rangle$ are taken from the Hilbert space $\mathcal{H}_{poly}= L^{2}(\mathbb{R}_{b},d\mu_{H})$, i.e. the set of square-integrable functions defined on the Bohr compactification of the real line $\mathbb{R}_{b}$ with a Haar measure $d\mu_{H}$. One chooses for them an inner product with a discrete normalization $\langle \nu |\mu \rangle = \delta_{\nu,\mu}$. The state of the system is described by a generic linear combination of them
\begin{equation}
|\psi\rangle =  \sum\limits_{i=1}^N a_{i}|\mu_{i}\rangle.
\end{equation}
One can identify two fundamental operators in this Hilbert space: a \textit{label operator} $\widehat{\varepsilon}$ and a \textit{shift operator} $\widehat{s}(\lambda)$. They act on the kets as follows
\begin{equation}
\label{label shift}
\widehat{\varepsilon} |\mu \rangle  = \mu |\mu \rangle \quad , \quad \widehat{s}(\lambda) |\mu \rangle = |\mu + \lambda \rangle.
\end{equation}
To characterize our system, described by the phase space variables $p$ and $q$, one assigns a discrete characterization to the variable $q$, and chooses to describe the wave function of the system in the so-called $p$-polarization. Consequently, the projection of the states on the pertinent basis vectors is
\begin{equation}
\phi_{\mu}(p) = \langle p | \mu \rangle = e^{-i\mu p}.
\end{equation}
Through the introduction of two unitary operators $U(\alpha)=e^{i\alpha \widehat{q}}, V(\beta) = e^{i\beta \widehat{p}} , (\alpha,\beta) \in  \mathbb{R}$ which obey the Weyl Commutation Rules (WCR) $U(\alpha)V(\beta) = e^{i\alpha \beta} V(\beta)U(\alpha)
$, one sees that the label operator is exactly the position operator, while it is not possible to define a (differential) momentum operator, as a consequence of the discontinuity for $\widehat{s}(\lambda)$ pointed out in Eq.(\ref{label shift}).
\subsection{The dynamical features}
\label{Dyn}
For the dynamical characterization of the model, the properties of the Hamiltonian system have to be investigated. The simplest Hamiltonian describing a one-dimensional particle of mass $m$ in a potential $V(q)$ is given by
\begin{equation}
\label{hamlib}
H = \frac{p^{2}}{2m} + V(q).
\end{equation}
In the $p$-polarization, as a consequence of the discreteness of $q$, it is not possible to define $\widehat{p}$ as a differential operator. 
The standard procedure is to define a subspace $ \mathcal{H}_{\gamma_{a}} $ of $ \mathcal{H}_{poly} $ containing all vectors that live on the lattice of points identified by the lattice spacing $a$
\begin{equation}
\gamma_{a} = \mathcal {f} q \in \mathbb {R} | q = na, \forall n \in \mathbb {Z} \mathcal {g},
\end{equation}
where $a$ has the dimensions of a \textit{length}.\\
Consequently, the basis vectors are of the form $ | \mu_ {n} \rangle $ (where $ \mu_ {n} = an $), and the states are all of the form
\begin{equation}
| \psi \rangle = \sum \limits_{n} b_{n} | \mu _ {n} \rangle.
\end{equation}
The basic realization of the polymer quantization is to approximate the term corresponding to the non-existent operator (this case $\widehat{p}$), and to find for this approximation an appropriate and well-defined quantum operator.
The operator $\widehat{V}$ is exactly the shift operator $\widehat{s}$, in both polarizations. Through this identification, it is possible to exploit the properties of $\widehat{s}$ to write an approximate version of $\widehat{p}$. For $p \ll \frac{1}{a}$, one gets
\begin{equation}
\label{appP}
p \simeq \frac{\sin(ap)}{a} = \frac{1}{2a}\left( e^{iap} - e^{-iap} \right)
\end{equation} 
and then the new version of $\widehat{p}$ is
\begin{equation}
\widehat{p}_{a} |\mu_{n} \rangle = \frac{i}{2a}\left( |\mu_{n-1}\rangle - |\mu_{n+1} \rangle \right).
\end{equation}
One can define an approximate version of $\widehat{p}^{2}$. For $p \ll \frac{1}{a}$, one gets
\begin{equation}
\label{appP2}
p^{2} \simeq \frac{2}{a^2} \left[ 1 - \cos (ap) \right] = \frac{2}{a^2} \left[1 - e^{iap} - e^{-iap} \right]
\end{equation}
and then the new version of $\widehat{p}^{2}$ is
\begin{equation}
\widehat{p}_{a}^{2} |\mu_{n} \rangle =  \frac{1}{a^2} \left[ 2|\mu_{n}\rangle - |\mu_{n+1}\rangle - |\mu_{n-1}\rangle \right].
\end{equation}
Remembering that $\widehat{q}$ is a well-defined operator as in the canonical way, the approximate version of the starting Hamiltonian (\ref{hamlib}) is
\begin{equation}
\label{Hpoly}
\widehat{H}_{a} = \frac{1}{2m}\widehat{p}_{a}^{2} + V(\widehat{q}).
\end{equation}
The hamiltonian operator $\widehat{H}_{a}$ is a well-defined and simmetric operator belonging to $ \mathcal{H}_{\gamma_{a}} $.
\subsection{The continuum Limit}
\label{Con}
The polymer representation of quantum mechanics is related with Schr\"odinger representation can now be analyzed.\\ 
Starting from a Hilbert space $\mathcal{H}_{poly}$, one needs to verify a limit operation to demonstrate that the space is isomorphic to the Hilbert space $\mathcal{H}_{S}=L^{2}(\mathbb{R},dq)$ \footnote{$L^{2}$ is the set of square-integrable functions defined on the real line $\mathbb{R}$ with a Lebesgue measure $dq$}.\\
The natural way to proceed is to start from a lattice $\gamma_{0}=\mathcal{f}q_{k}\in \mathbb{R}| q_{k}=ka_{0}, \forall k\in \mathbb{Z}\mathcal{g}$ and subdivide each interval $a_{0}$ in $2^{n}$ intervals of length $a_{n}=\frac{a_{0}}{2^{n}}$. Unfortunately, this is not possible because, when densifying the lattice, the elements of $\mathcal{H}_{poly}$ have a norm that tend to infinity. This is because $\mathcal{H}_{S}$ and its states cannot be included in $\mathcal{H}_{poly}$. However, it is possible to realize a different procedure. From a continuous wave function, one has to find the best wave function defined on the lattice that approximates it, in the limit when the lattice becomes denser. The strategy to properly implement this approach is the introduction of a scale $C_{n}$, which, in our case, is the subdivision of the real line into disjoint intervals of the form $\alpha_{i}=[ia_{n},(i+1)a_{n})$, where the extrema of the range are the lattice points. On this level, one approximates continuous functions with constant intermediate states belonging to $\mathcal{H}_{S}$. 
So for, one has a whole series of effective theories, depending on the scale $C_{n}$, that approximate much and much better the continuous functions and that have a well defined Hamiltonian. As in \cite{corichidue}, by introducing a cut-off for each Hamiltonian defined on the intervals, making the operation of coarse graining and entering a normalization factor in the internal product, one verifies that the existence of continous limit is equivalent to the description of the energy spectrum (relative to the Hamiltonian defined after the cut-off) as tending to the continuous spectrum, such that a complete set of normalized eigenfunctions exists.
The space obtained this way is isomorphic to the space $ \mathcal{H}_{S}$.
\subsection{The Free Polymer particle}
\label{FPP}
In this sub-section, we analize the simplest one-dimensional system in the presence of a discrete structure of the space variable $q$, i.e. the free polymer particle\cite{taub}. When the free polymer particle problem is taken into account, the potential term in Eq.(\ref{Hpoly}) is negligible. Therefore, in the $p$-polarization, the quantum state of the system is described by the wave function $\psi(p)$ via the eigenvalue problem
\begin{equation}
\label{partlibpol}
\left[ \frac{1}{ma^{2}} \left( 1-\cos (ap) \right) - E_{a}\right]\psi(p) = 0.
\end{equation}
Here, $E_{a}$ is an eigenvalue depending on the scale $a$, and one has 
\begin{equation}
\label{spe par lib}
E_{a} = \frac{1}{ma^{2}} \left[ 1 - \cos (ap) \right] \leq \frac{2}{ma^{2}} = E_{a}^{max}.
\end{equation}
From Eq.(\ref{spe par lib}), one sees that, for each scale $a$, there is a bounded and continous eigenvalue. In the limit $a\rightarrow0$, i.e. \textit{switching} the polymer effect off, one obtains the unbounded eigenvalue $E = \frac{p^{2}}{2m}$, typical for a free particle.
It is easy to verify that the solution $\psi(p)$ for the eigenvalue problem (\ref{partlibpol} has the form
\begin{equation}
\label{autfunzfree}
\psi(p) = A\delta(p - P_{a}) + B\delta(p + P_{a}),
\end{equation}
where $A,B$ are integration constants and 
\begin{equation}
\label{reldismod}
P_{a} = \frac{1}{a}\arccos ( 1 - m a^{2}E_{a} )
\end{equation}
induces modified dispersion relation in the presence of a polymer structure.\\
For an (inverse) Fourier transform for the eigenfunction (\ref{autfunzfree}), one obtains the eigenfunction in the $q$-polarization $\psi(q)$ as
\begin{equation}
\label{autofunz lib}
\psi(q) = \int \psi(p) e^{ipq} = A e^{iqP_{a}} + B e^{-iqP_{a}}.
\end{equation}
The eigenfunction in the $q$-polarization becomes a modified wave plane, due to the dispersion relation (\ref{reldismod}), which are valid at each scale.
\subsection{The polymer particle in a box}
\label{PPB}
In this subsection, we will analyze the dynamical features of a one-dimensional particle in a box, within the framework of the polymer representation of quantum mechanics. 
For a one-dimensional box (i.e. a segment) of length $L=na, n \in \mathbb{N}$, the potential $V(q) = V(na)$ reads
\begin{equation}
\label{potbox}
V(q) = \begin{cases} \infty, & x>L , x<0  \\ 0, & 0<x<L \end{cases} ,
\end{equation}
i.e. in the case of a potential limited by infinite walls. In this case, the particle behaves as a free particle within the segment, and proper boundary conditions for eigenfunction (\ref{autofunz lib}) have to be imposed. In particular
\begin{equation}
\label{cond part sca}
\psi(0) = \psi(L) = 0 \longrightarrow \begin{cases}& A = -B \\ & LP_{a} = n\pi \end{cases},
\end{equation}
for which the eigenfunctions $\psi(q)$ in the $q$ polarization are obtained
\begin{equation}
\psi(q) = 2A \sin \left(\frac{n\pi q}{L} \right).
\end{equation}
The corresponding energy spectrum $E_{a,n}$ is a function of both the lattice constant $a$ and the quantum number $n$, such that
\begin{equation}
\label{spettro poly}
E_{a,n} = \frac{1}{m a^{2}}\left[ 1 - \cos \left( \frac{a n \pi}{L} \right) \right].
\end{equation}
In the limit $a\rightarrow0$ one gets the energy spectrum of the standard case.
\section{THE MIXMASTER MODEL: CLASSICAL AND QUANTUM FEATURES}
\label{MIX}
In this section, we provide a complete description of the most relevant achievements obtained for the dynamics of the Bianchi IX cosmological model, both in the classical and the quantum regime towards the cosmological singularity, as they are depicted in the two pioneering works\cite{misnermixmaster},\cite{misner}.
\subsection{The classical dynamics}
\label{Cla}
Homogeneous spaces are an important class of cosmological models. These spaces are characterized by the preservation of the space line element under a specific group of symmetry, and are collected in the so-called Bianchi classification\cite{landau}. The most general homogeneous model is the Bianchi IX model. As demonstrated by Belinski, Khalatnikov and Lifshitz (BKL)\cite{cinque}, when a generic inhomogeneous space approaches the singularity, it behaves as an ensemble of Bianchi IX independent models in each point of space\footnote{Also the Bianchi VIII as the same degree of generality but it does not admit an isotropic limit}. 
Following the Misner parametrization\cite{misnermixmaster}, the line element for the Bianchi IX model is
\begin{equation}
\label{pappa}
ds^{2} = N(t)^{2}dt^{2} - \eta_{ab}\omega^{a}\omega^{b},
\end{equation}
where 
$\omega ^{a} = \omega^{a}_{\alpha}dx^{\alpha}$ is a set of three invariant differential forms, $N(t)$ is the lapse function and $\eta _{ab}$ is defined as $\eta _{ab} = e^{2\alpha}(e^{2\beta})_{ab}$. 
In the Misner picture, $\alpha$ expresses the isotropic volume of the universe (for $\alpha\rightarrow -\infty$, the initial singularity is reached.), while the matrix $\beta_{ab} = diag(\beta _{+} + \sqrt{3}\beta _{-} , \beta _{+} - \sqrt{3}\beta _{-} , -2\beta _{+})$ accounts for the anisotropy of this model. The introduction of the Misner variables allows one to rewrite the super Hamiltonian constraint (written following the ADM formalism\cite{ADM}) in this simple way
\begin{equation}
\label{vincolo misner}
\mathcal{H}_{IX} = -p_{\alpha}^{2}+p_{+}^{2}+p_{-}^{2}+\frac{3(4\pi)^{4}}{k^{2}}e^{4\alpha}V(\beta _{\pm}) = 0,
\end{equation}
where the ($p_{\alpha},p_{\pm}$) are the conjugated momenta to ($\alpha,\beta_{\pm}$) respectively, $k = 8 \pi G$, and $V(\beta _{\pm})$ is the potential term depending only on $\beta_{\pm}$, i.e. the anisotropies.
\begin{multline}
\label{potanis}
 V(\beta _{\pm})= e^{-8\beta _{+}}-4e^{-2\beta _{+}}\cosh (2\sqrt{3}\beta _{-})+ \\ +2e^{4\beta _{+}}\left[ \cosh(4\sqrt{3}\beta_{-})-1 \right].
\end{multline}
\begin{figure}[h!]
\centering
\includegraphics[scale=.35]{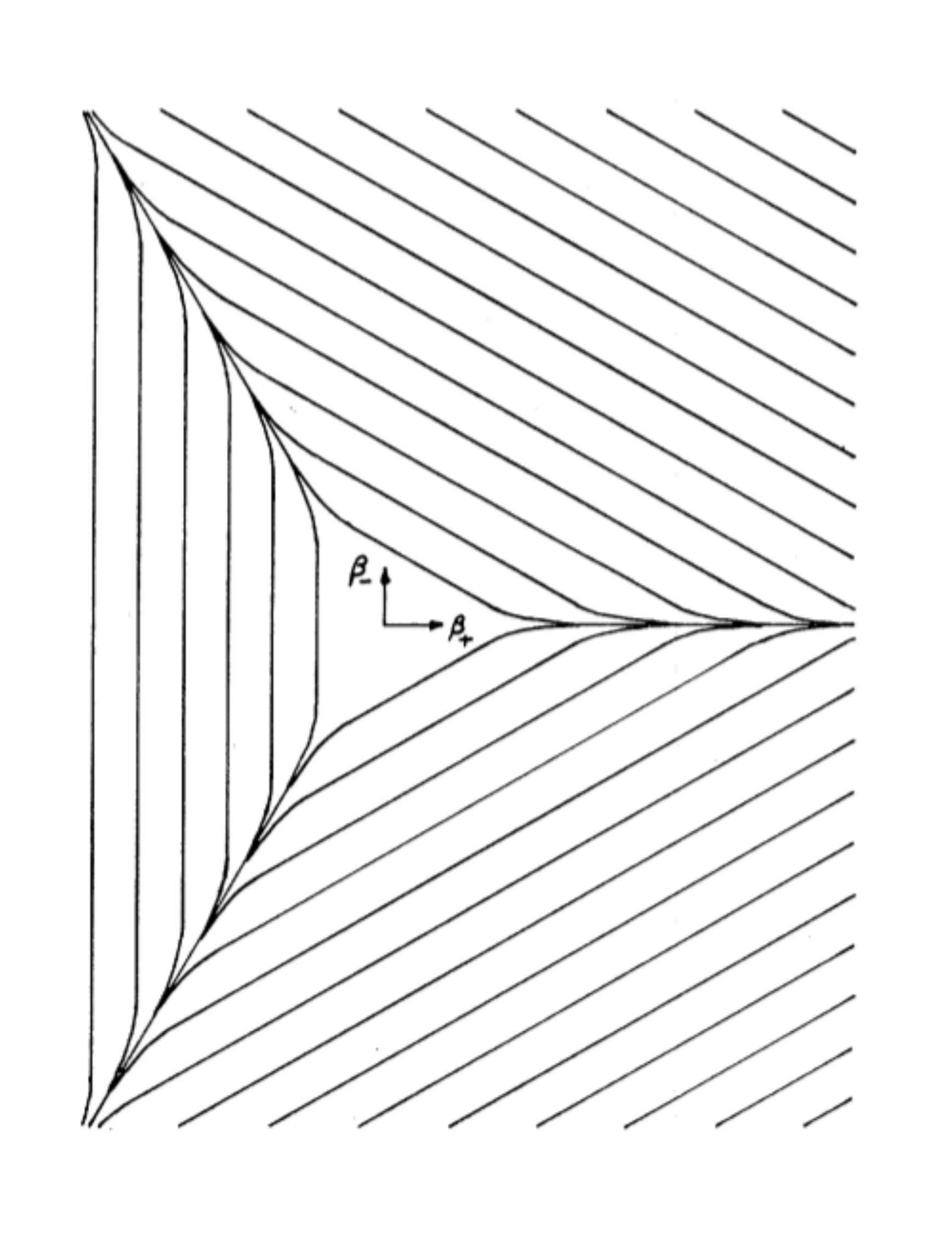}
\caption{ \footnotesize "(Color online)".Equipotential lines of Bianchi IX model in ($\beta_{+},\beta_{-}$) plane\cite{misner}.}
\label{fig:potBIX}
\end{figure}
Let us execute now the ADM reduction of the dynamics\cite{ADM2} by solving the super Hamiltonian constraint with respect to a specific conjugated momenta and then by identifing a time-variable for the phase space. For the purposes of this investigation, we choose to solve (\ref{vincolo misner}) with respect to $p_{\alpha}$ and identify $\alpha$ as a time-variable. This choice is justified because, if we choose a time gauge $\dot{\alpha}=1$ in the synchronous reference system ($N(t)=1$), the isotropic volume $\alpha$ depends on the synchronous time $t$ by the relation $\alpha=\frac{1}{3}\ln t$. Then one obtains
\begin{equation}
\label{HBIXmisner}
-p_{\alpha} = \mathcal{H}_{ADM} \equiv \sqrt{p_{+}^{2}+p_{-}^{2}+\frac{3(4\pi)^{4}}{k^{2}}e^{4\alpha}V(\beta _{\pm})} ,
\end{equation}
i.e. the so-called \textit{reduced Hamiltonian} of our problem. 
From relation (\ref{HBIXmisner}), one recognizes that, as studied by C.W.Misner\cite{misner}, the dynamics of the Universe towards the singularity is mapped to the description of the
motion of a particle that lives on a plane inside a closed domain.
This way, we can study how the anisotropies $\beta_{\pm}$ change with respect to the time variable $\alpha$ through equations of motion related to the reduced Hamiltonian
\begin{equation}
\begin{split}
\label{eqHam}
&\beta '_{\pm}=\frac{d \beta _{\pm}}{d \alpha} = \frac{p_{\pm}}{\mathcal{H}_{ADM}},\\
&p_{\pm} '=\frac{d p_{\pm}}{d \alpha} = \frac{3(4\pi)^{4}}{2k\mathcal{H}_{ADM}}e^{4\alpha}\frac{\partial V(\beta _{\pm})}{\partial \beta _{\pm}}.
\end{split}
\end{equation}
Studying the two opposite approximations of $V(\beta_{\pm})$, i.e. far from the walls ($V\simeq0$) and close to the walls ($V\simeq\frac{1}{3}e^{-8\beta_{+}}$), we can obtain the relative motion between the particle and the potential wall. It is possible to obtain, for $V\simeq0$, the behavior of $\beta_{\pm}$ as a function of time $\alpha$ via a simple integration of the first equation of motion. This way, one gets
\begin{equation}
\label{andamentoMis}
\beta_{\pm}\propto \frac{p_{\pm}}{\sqrt{p_{+}^{2} + p_{-}^{2}}}\alpha.
\end{equation}
Moreover, the \textit{anisotropy velocity} of the particle far from the walls is defined as
\begin{equation}
\label{velaniso}
\beta ^{'} = \sqrt{\left(\frac{d\beta _{+}}{d\alpha}\right)^{2}+\left(\frac{d\beta _{-}}{d\alpha}\right)^{2}} = 1
\end{equation}
for each value of $p_{\pm}$.
On the other hand, the investigation on the motion of one of the equivalent sides allow one to understand that the walls move towards the 'outer' directionou with velocity $|\beta'_{w}|=\frac{1}{2}$. 
The particle always collides against the wall and bounces from one to another.
This chaotic dynamics is the analogue of the oscillatory regime described by BKL in \cite{tre}.
\newline
It is worth noting that the regime under which $V\simeq0$ corresponds to the Bianchi I case of the Bianchi classification, the so-called \textit{Kasner regime}, in which the particle moves as being free and the two constraints 
\begin{equation}
\begin{split}
\label{somma indici}
&p_{1}+p_{2}+p_{3} = 1, \\
&p_{1}^{2}+p_{2}^{2}+p_{3}^{2} = 1,
\end{split}
\end{equation}
are satisfied. Here $p_{1},p_{2},p_{3}$ are the \textit{Kasner indices}, i.e. three real numbers that express the anisotropy of the model. Writing the spatial metric of the Bianchi I model in the synchronous reference system, i.e.
\begin{equation}
\label{metricaBI}
dl^{2}=t^{2p_{1}}dx^{1}+t^{2p_{2}}dx^{2}+t^{2p_{3}}dx^{3},
\end{equation}
the presence of the Kasner indices inside the spatial metric is understood to imply a different behavior along the different directions, which define the anisotropic directions.\\ 
On the other hand, when the particle is close the wall ($V\simeq\frac{1}{3}e^{-8\beta_{+}}$), the Bianchi II model, i.e. the model descrbing one single bounce against infinite wall potential\cite{montanireview}, is considered. 
The system (\ref{eqHam}) can be studied close to a potential wall, and is possible to identify two constants of motion
\begin{equation}
\begin{split}
\label{pmeno}
&p_{-} = cost, \\
&K = \frac{1}{2}p_{+} + \mathcal{H}_{ADM} = cost.
\end{split}
\end{equation}
These relations have been obtained for the 'vertical' potential wall in Fig.(\ref{fig:potBIX}); it is however necessary to stress that the bounces against the potential walls are all equivalent as far as the dynamics of the system is concerned, as the potential walls can be obtained one from the other, by taking into account the symmetries of the model, as analyzed in \cite{indicikasner}.\\
A description of this regime is illustrated in Fig.(\ref{fig:potBIX}).
The anisotropies can be parameterized as functions of both the incidence angle and of the reflection one, $\theta_{i}$ and $\theta_{f}$, respectively. This way,
\begin{equation}\begin{split}
\label{param}
&(\beta_{-}')_{i} = \sin\theta_{i},\\
&(\beta_{+}')_{i} = -\cos\theta_{i},\\
&(\beta_{-}')_{f} =\sin\theta_{f},\\
&(\beta_{+}')_{f} = \cos\theta_{f}.
\end{split}\end{equation}
The relations (\ref{pmeno}) are used to obtain a reflection law for a generic single bounce
\begin{equation}
\label{mappamisner}
\sin\theta_{f} - \sin\theta_{i} = \frac{1}{2}\sin(\theta_{i}+\theta_{f}).
\end{equation}
However, there is a maximum angle $\theta_{max}$ after which no bounce occurs. For the occurrence of a bounce, the longitudinal component of the velocity $\beta'_{+}$ must be greater than the wall velocity $\beta'_{w}$. This condition is expressed as
\begin{equation}
\label{angolo}
|\theta_{i}| < |\theta_{max}| = \arccos\left( \frac{\beta'_{w}}{\beta'_{+}}\right) = \frac{\pi}{3}.
\end{equation}
As a result, the particle,sooner or later, will assume all the possible directions, regardless of the initial condition. Following the convenience choice used by C.W. Misner in \cite{misner}, and taking advantage of the geometric properties of this scheme, in the limit close to the singularity ($\alpha \rightarrow -\infty$) one finds a conservation law of the form
\begin{equation}
\label{conservazione}
< \mathcal{H}_{ADM}\alpha > = cost.
\end{equation} 
For two successive bounces (the $i$-th and the $(i+1)$-th of the sequence), $\alpha^{i}$ expresses the time at which the $i$-th bounce occurs and $\mathcal{H}_{ADM}^{i}$ the value of reduced Hamiltonian (\ref{HBIXmisner}) just before the $i$-th bounce: relation (\ref{conservazione}) states that
\begin{equation}
\label{bounce}
\mathcal{H}_{ADM}^{i}\alpha^{i} = \mathcal{H}_{ADM}^{i+1}\alpha^{i+1}.
\end{equation}
In other words, the quantity $\mathcal{H}_{ADM}\alpha$ acquires the same costant value as just before each bounce towards the singularity.
\subsection{The quantum behavior}
\label{Qua}
The canonical quantization of the system consists of the commutation relations
\begin{equation}
[\widehat{q}_{a},\widehat{p}_{b}] = i\delta _{ab},
\end{equation}
which are satisfied for $\widehat{p_{a}} = -i\frac{\partial}{\partial q_{a}}=-i\partial_{a}$ where $a,b=\alpha, \beta_{+},\beta_{-}$.
By replacing the canonical variables with the corresponding operators, the quantum behavior of the Universe is given by the quantum version of the superhamiltonian constrain (\ref{vincolo misner}), i.e. the \textit{Wheeler-deWitt equation}(WDW) for the Bianchi IX model
\begin{multline}
\label{WDWstandard}
\widehat{\mathcal{H}}_{IX}\Psi(\alpha,\beta_{\pm}) = \\ = \left[\partial_{\alpha}^{2} - \partial_{+}^{2} - \partial_{+}^{2} + \frac{3(4\pi)^{4}}{k^{2}}e^{4\alpha}V(\beta _{\pm})\right]\Psi(\alpha,\beta_{\pm}),
\end{multline}
where $\Psi(\alpha,\beta_{\pm})$ is the wave function of the Universe which provides information about the physical state of the Universe.
A solution of Eq.(\ref{WDWstandard}) can be looked for in the form
\begin{equation}
\label{funzuni}
\Psi = \sum _{n} \chi _{n}(\alpha)\phi_{n}(\alpha,\beta).
\end{equation}
The \textit{adiabatic approximation} consists in requiring that the $\alpha$-evolution be principally contained in the $\chi _{n}(\alpha)$ coefficients, while the functions $\phi_{n}(\alpha,\beta)$ depend on $\alpha$ parametrically only. The adiabatic approximation is therefore expressed by the condition
\begin{equation}
\label{adiabatic}
|\partial_{\alpha}\chi _{n}(\alpha)| \gg |\partial_{\alpha}\phi_{n}(\alpha,\beta)|.
\end{equation}
By applying condition (\ref{adiabatic}), the WDW Eq.(\ref{WDWstandard}) reduces to an eigenvalue problem related to the reduced hamiltonian $\mathcal{H}_{ADM}$ via
\begin{multline}
\label{H2quantum}
\widehat{\mathcal{H}}_{ADM}^{2}\phi_{n} = E^{2}_{n}(\alpha)\phi_{n} = \\ =\left[ -\partial_{+}^{2} - \partial_{-}^{2} + \frac{3(4\pi)^{4}}{k^{2}}e^{4\alpha}V(\widehat{\beta _{\pm}})\right]\phi_{n}.
\end{multline}
However, even without finding the exact expression of the eigenfunctions, one may gain important information about the system from a quantum point of view near the initial singularity.
From Fig.(\ref{fig:potBIX}), one can see how the potential (\ref{potanis}) can be modelized as an infinitely steep potential well with a triangular base. In \cite{misner}, the strong hypothesis to replace the triangular box with a squared box having the same area $L^{2}$ is proposed.  
This way, the probelm describing a two-dimensional particle in a squared box with infinite walls is recovered. In this case, the eigenvalue problem becomes
\begin{equation}
\label{boxsquare}
\widehat{\mathcal{H}}_{ADM}^{2}\phi_{n,m} = \frac{\pi ^{2}(m^{2}+n^{2})}{L^{2}(\alpha)}\phi_{n,m},
\end{equation}
where $m,n\in\mathbb{N}$ are the quantum numbers associated to ($\beta_{+},\beta_{-}$).
By a direct calculation, we can derive $L^{2}(\alpha)=\frac{3\sqrt{3}}{4}\alpha ^{2}$, such that the eigenvalue is
\begin{equation}
\label{autovaloremisner}
E_{n,m}=\frac{2\pi}{3^{3/4}\alpha}\sqrt{m^{2}+n^{2}}.
\end{equation}
As demonstrated in \cite{soluzione}, substituting the eigenvalue expression (\ref{autovaloremisner}) in the Eq.(\ref{WDWstandard}), the self-consistence of adiabatic approximation is ensured.
Let us use (\ref{autovaloremisner}) with (\ref{conservazione}) to estimate the quantum numbers behavior towards the singularity. One can see in Eq.(\ref{autovaloremisner}) that the eigenvalue spectrum is unlimited from above, such that, for sufficiently high occupation numbers, the replacing $\mathcal{H}_{ADM} \simeq E_{n,m}$ is a good approximation. This way, for $\alpha\rightarrow-\infty$, Eq.(\ref{conservazione}) becomes
\begin{equation}
\label{numerioccupazione}
<\mathcal{H}_{ADM}\alpha >  \xrightarrow [\alpha \rightarrow - \infty]{} < \sqrt{m^{2}+n^{2}} > = cost.
\end{equation}
Being the current state of the Universe anisotropy characterized by a classical nature, i.e. $\sqrt{m ^ {2} + n ^ {2}} >> 1 $, we can say, by Eq.(\ref{numerioccupazione}), that this quantity is constant approaching the singularity. This way, the quantum state of the Universe related to the anisotropies remains classical for all the backwards history until the singularity.
\section{ SEMICLASSICAL Polymer approach to the MIXMASTER MODEL}
\label{SEMI}
The aim of the present Section is to discuss how to apply the \textit{polymer} approach of Sec.\ref{PQM} to the Bianchi IX model at a semiclassical level and to verify if and how the nature of the cosmological singularity is modified. Here, ``semiclassical'' means that we are working with a modified super Hamiltonian constraint obtained as the lowest order term of a WKB expansion for $\hbar\rightarrow0$. At this level, the modified theory is subject to a deterministic dynamics. Following the procedure in Sec.\ref{Dyn}, one can choose, with a precise physical interpretation, to define the anisotropies of the Universe  $(\beta_{+},\beta_{-})$ as discrete variables leaving the characterization of the isotropic variable $\alpha$ unchanged, which here plays the role of time. 
This procedure formally consists in the replacement
\begin{equation}
\label{subpol}
p_{\pm}^{2}\rightarrow \frac{2}{a^{2}}\left[1-\cos (a p_{\pm})\right].
\end{equation}
The superhamiltonian constraint (\ref{vincolo misner}) becomes
\begin{equation}
\label{vinSHpoly}
-p_{\alpha}^{2}+\frac{2}{a^{2}}\left[2-\cos (a p_{+})-\cos (a p_{-})\right]+\frac{3(4\pi)^{4}e^{4\alpha}}{k^{2}}V(\beta _{\pm}) = 0.
\end{equation}
We define $-p_{\alpha} \equiv H_{poly}$ as the \textit{reduced Hamiltonian}, such that one gets
\begin{multline}
\label{ADMpoly}
-p_{\alpha} \equiv H_{poly}= \\ =\sqrt{ \frac{2}{a^{2}}\left[2 - \cos(a p_{+}) - \cos(a p_{-})\right] + \frac{3(4\pi)^4e^{4\alpha}}{k^2}V(\beta_{\pm})}.
\end{multline}
Starting from the new hamiltonian formulation (\ref{ADMpoly}), we can get the following set of the hamiltonian equations as
\begin{equation}
\begin{split}
\label{eqHampoly}
&\beta '_{\pm}=\frac{d \beta _{\pm}}{d \alpha} =  \frac{\sin (a p_{\pm})}{a H_{poly}},\\
&p_{\pm} '=\frac{d p_{\pm}}{d \alpha} = \frac{3(4\pi)^{4}}{2kH_{poly}}e^{4\alpha}\frac{\partial V(\beta _{\pm})}{\partial \beta _{\pm}}.
\end{split}
\end{equation}
This modification leaves the potential $V(\beta_{\pm})$ and the isotropic variable $\alpha$ unchanged. Therefore, even in the modified theory, the walls move in the 'outer' direction with velocity $|\beta'_{w}|=\frac{1}{2}$ and the initial singularity is not expected to be removed.
\\
Let us start by analyzing the system far from the wall, i.e. with $V\simeq0$. As one can see in (\ref{eqHampoly}) when $V\simeq0$, the anisotropy velocity is modified if compared to the standard case. 
In particular, the behavior of $\beta_{\pm}$ is proportional to the time $\alpha$, as in the standard theory, but with a different coefficient, i.e.
\begin{equation}
\label{andamentoPoly}
\beta_{\pm}\propto \frac{\sin(ap_{\pm})}{\sqrt{4 - 2[\cos(ap_{+}) + \cos(ap_{-})]}}\alpha.
\end{equation}
In particular, by the definition of the anisotropy velocity, Eq.(\ref{velaniso}), one obtains
\begin{equation}
\label{velpart}
\beta' = \sqrt{\frac{\sin(ap_{+})^{2} +\sin(ap_{-})^{2}}{4 - 2[\cos(ap_{+}) + \cos(ap_{-})]}} = r(a,p_{\pm}).
\end{equation}
It is worth noting that $r(a,p_{\pm})$ is a bounded function ($r\in[0,1]$) of parameters that remains constant between one bounce and the following one. From Eq.(\ref{andamentoPoly}), we have a Bianchi I model modified by the \textit{polymer} substitution. As a consequence of this feature, also in the modified theory, the anisotropies behaves respect to $\alpha$ in a proportional way.
The first important semiclassical result is the relative motion between wall and particle. From (\ref{velpart}), one can observe the existence of allowed values of $(ap_{+},ap_{-})$, such that the particle velocity is smaller than the wall velocity $\beta'_{w}$. Therefore, the condition for a bounce is
\begin{equation}
\label{cond}
\beta' = \sqrt{\frac{\sin(ap_{+})^{2} +\sin(ap_{-})^{2}}{4 - 2[\cos(ap_{+}) + \cos(ap_{-})]}} > \frac{1}{2} =\beta'_{w}.
\end{equation}
It means that the infinite sequence of bounces against the walls, typical of the Mixmaster Model, takes place until condition (\ref{cond}) is valid. When $r<\frac{1}{2}$, the particle becomes slower than the potential wall and reaches the singularity without no other bounces.
\\
This feature is confirmed by the analysis on the Kasner relations (\ref{somma indici}). The first Kasner relation is still valid in the deformed approach, because the sum of the Kasner indices is linked by the Misner variables just trough the isotropic variable $\alpha$. Instead, the second Kasner relation is directly related to the anisotropy velocity \cite{montanireview} and it results modified into
\begin{multline}
\label{relmod}
p_{1}^{2}+p_{2}^{2}+p_{3}^{2} = \frac{1}{3} + \frac{2}{3} \left[ (\beta_{+}')^{2} + (\beta_{-}')^{2} \right] = \\ = 1 - \frac{2}{3}(1-r^{2}) = 1 - q^{2},
\end{multline}
where $q^{2}=\frac{2}{3}(1-r^{2})$.
We introduce $q^{2}$ in (\ref{relmod}) because $\frac{2}{3}(1-r^{2})\geq0$ for any values of $(ap_{+},ap_{-})$. The introduction of the \textit{polymer} structure for the anisotropies acts the same way as a massless scalar field in Bianchi IX model\cite{berger},\cite{intersezione}. For this reason, we can choose to describe the Kasner indices with the same parametrization due to Belinski and Khalatnikov in \cite{scalar field}. It is realized through the introduction of two parameters $(u,q)$\footnote{In the standard case (absence of the scalar field or \textit{polymer} modification, i.e. $q^{2}=0$) $p_{1},p_{2},p_{3}$ and $u$ are related this way\cite{indicikasner}:\\ $p_{1}(u) = -\frac{u}{1+u+u^{2}} , p_{2}(u) = \frac{1+u}{1+u+u^{2}} , p_{3}(u) = \frac{u(1+u)}{1+u+u^{2}}$.\\ In this case $0<u<1$.} and it allows to represent the all possible values of the Kasner indices. One gets
\begin{equation}
\begin{split}
\label{paramscala}
&p_{1} = \frac{-u}{1+u+u^{2}},\\
&p_{2} = \frac{1+u}{1+u+u^{2}} \left[ u - \frac{u-1}{2}(1-\sqrt{1 - \gamma^{2}}) \right],\\
&p_{3} = \frac{1+u}{1+u+u^{2}} \left[ 1 + \frac{u-1}{2}(1-\sqrt{1 - \gamma^{2}}) \right],\\
&\gamma^{2} = \frac{2(1+u+u^{2})q^{2}}{(u^{2}-1)^{2}}.
\end{split}\end{equation}
Here, $-1<u<+1$ and $-\sqrt{\frac{2}{3}}<q<\sqrt{\frac{2}{3}}$.
The presence of $\gamma^{2}$ inside Eq.'s(\ref{paramscala}) means that not all values of $u,q$ are allowed. The $u,q$ allowed values are those which respect the condition $\gamma^{2}<1$.
Inside this region of permitted values, there are two fundamental areas where all Kasner indices are simultaneously positive, i.e. for $q>\frac{1}{\sqrt{2}}$ and $q<-\frac{1}{\sqrt{2}}$.
When it happens, remembering that the spatial Kasner metric is $dl^{2}=t^{2p_{1}}dx^{1}+t^{2p_{2}}dx^{2}+t^{2p_{3}}dx^{3}$, the distances contract along all the spatial direction approaching the singularity ($t\rightarrow0$). It means that the system behaves as a stable Kasner regime and the oscillatory regime is suppressed.
\\
Furthemore, the relation (\ref{bounce}) remains valid until $r<\frac{1}{2}$ or rather when the particle become slower than the potential wall. When it happens, approching the singularity, $\alpha\rightarrow -\infty$ while $H_{poly}$ remains costant without changes. In this sense, when the outgoing momenta configuration of the $j$-th bounce is such that $r<\frac{1}{2}$, the quantity $H_{poly}^{j}\alpha^{j}$ is no longer a constant of motion.
\\
As in the standard case, we can introduce a parametrization for the particle velocity components, before and after a single bounce
\begin{equation}
\begin{split}
\label{parpoly}
&(\beta_{-}')_{i} = r_{i}\sin\theta_{i},\\
&(\beta_{+}')_{i} = -r_{i}\cos\theta_{i},\\
&(\beta_{-}')_{f} = r_{f}\sin\theta_{f},\\
&(\beta_{+}')_{f} = r_{f}\cos\theta_{f}.
\end{split}
\end{equation}
where $(\theta_{i},\theta_{f})$ are the incidence and the reflection angles and $(r_{i},r_{f})$ are the anisotropy velocities before and after the bounce.
Eq. (\ref{angolo}) states the existence of a maximum angle $\theta_{max}=\frac{\pi}{3}$ for a bounce to occur.\\
In the modified model, the condition for a bounce to take place is 
\begin{equation}
\label{inclinazione poly}
\theta_{i} < \theta_{max}^{poly} = \arccos(\frac{1}{2r_{i}}) \leq \arccos(\frac{1}{2}) = \theta_{max} = \frac{\pi}{3}.
\end{equation}
\begin{figure}[h!]
\centering
\includegraphics[scale=.67]{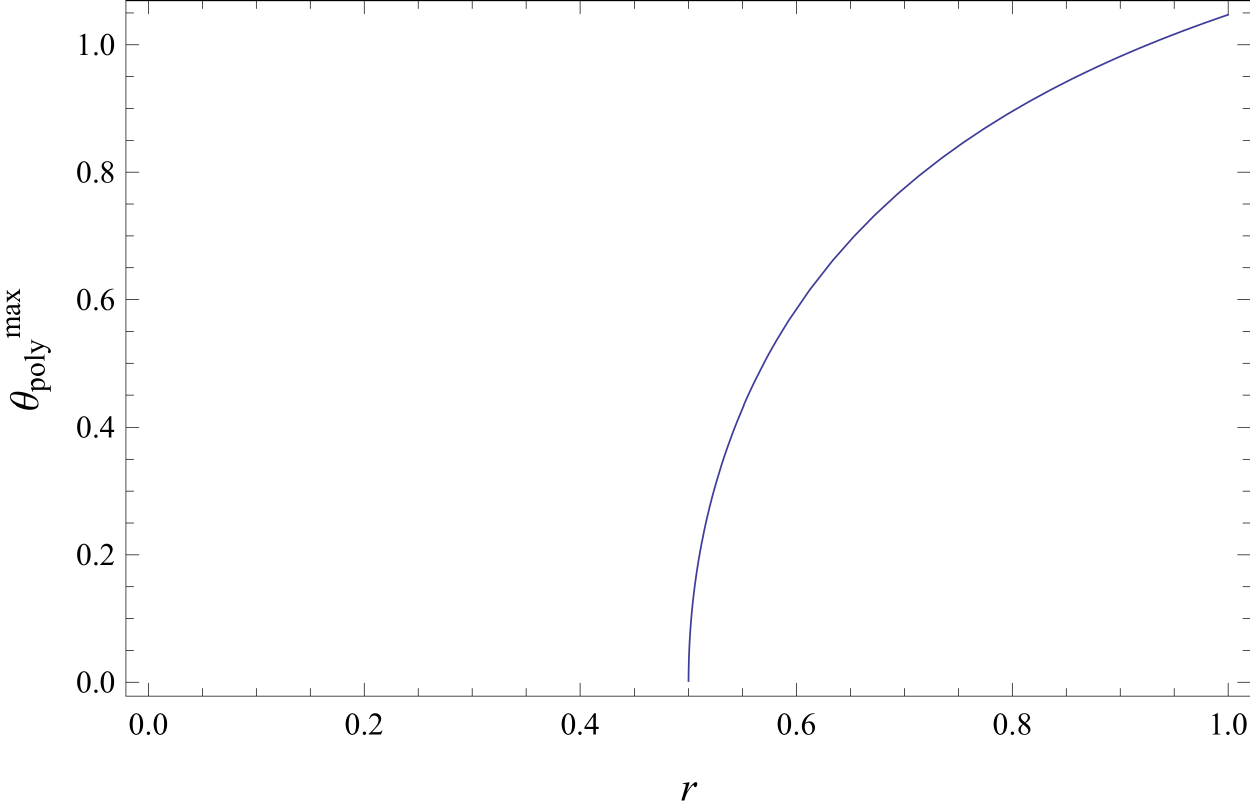}
\caption{\footnotesize "(Color online)".The maximum angle for have a bounce $\theta_{max}^{poly}$ as a function of $r$. In the $r\rightarrow1$ limit, the standard case is restored. This treatment make sense only for a configuration in which the particle velocity is higher than walls velocity, i.e. for $r>\frac{1}{2}$.}
\label{fig:inc poly}
\end{figure}
The new maximum angle $\theta_{max}^{poly}$ coincides with $\theta_{max}$ just for $r=1$, i.e. when the standard case is restored (Fig. (\ref{fig:inc poly})).
The last semiclassical result is the modified reflection law for a single bounce: as in the standard case, we can identify two constants of motion by studying the system near the potential wall. In particular, one has
\begin{equation}
\begin{split}
\label{kappapol}
&p_{-} = cost, \\
&K = \frac{1}{2}p_{+} + H_{poly} = cost.
\end{split}
\end{equation}
The expression of $p_{+}$ as function of $\beta^{'}$ can be obtained from (\ref{eqHampoly}):
\begin{equation}
\label{pinfunzbeta}
p_{+} = \frac{1}{a}\arcsin(a\beta_{+}'H_{poly}).
\end{equation}
This way, by a substitution of Eq.(\ref{pinfunzbeta}) in Eq.(\ref{kappapol}), remembering $\arcsin(-x) = -\arcsin(x)$ and using the parametrization (\ref{parpoly}), one obtains
\begin{multline}
\frac{1}{2a}\arcsin(-ar_{i}H_{poly}^{i}\cos\theta_{i}) + H_{poly}^{i} = \\ =\frac{1}{2a}\arcsin(ar_{f}H_{poly}^{f}\cos\theta_{f}) + H_{poly}^{f}.
\end{multline}
Now we express $r$ and $H_{poly}$ as functions of $a,p_{+},p_{-}$:
\begin{multline}
\label{dasviluppare}
\frac{1}{2}[\arcsin(\sqrt{\sin(ap_{+}^{i})^{2} +\sin(ap_{-}^{i})^{2}} \cos\theta_{i}) + \\ +\arcsin(\sqrt{\sin(ap_{+}^{i})^{2} +\sin(ap_{-}^{i})^{2}}\frac{\cos\theta_{f}\sin\theta_{i}}{\sin\theta_{f}})] = \\
= \sqrt{4 - 2(\cos(ap_{+}^{i}) + \cos(ap_{-}^{i})} - \frac{\sin\theta_{i}}{\sin\theta_{f}} \times \\ \times \sqrt{\frac{\sin(ap_{+}^{i})^{2} +\sin(ap_{-}^{i})^{2}} {\sin(ap_{+}^{f})^{2} +\sin(ap_{-}^{f})^{2}}[4 - 2(\cos(ap_{+}^{f} + \cos(ap_{-}^{f})] } .
\end{multline}
To perform a direct comparison with the standard case, a Taylor expansion up to second order for $ap_{\pm}<<1$ for Eq.(\ref{dasviluppare}) is needed. This way, after standard manipulation, the reflection law rewrites
\begin{multline}
\frac{1}{2}\sin(\theta_{i} + \theta_{f}) = \sin\theta_{f}\sqrt{1 + \frac{a^{2}}{4}  \frac{(p_{+}^{i})^{4}+(p_{-}^{i})^{4}}{(p_{+}^{i})^{2}+(p_{-}^{i})^{2}} } - \\ - \sin\theta_{i}\sqrt{1 + \frac{a^{2}}{4} \frac{(p_{+}^{f})^{4}+(p_{-}^{f})^{4}}{(p_{+}^{f})^{2}+(p_{-}^{f})^{2}}}.
\end{multline}
Defining $R=\frac{a^{2}}{4}  \frac{p_{+}^{4}+p_{-}^{4}}{p_{+}^{2}+p_{-}^{2}}$, one has
\begin{equation}
\frac{1}{2}\sin(\theta_{i} + \theta_{f}) = \sin\theta_{f}\sqrt{1 + R_{i}} - \sin\theta_{i}\sqrt{1 + R_{f}}.
\end{equation}
We obtain for $ap_{\pm}<<1$ a modified reflection law that, differently from the standard case, depends on two parameters $(R,\theta)$. Obviously, in the limit $ap_{\pm} \rightarrow 0$, i.e. switching off the \textit{polymer} modification, the standard reflection law (\ref{mappamisner}) is recovered.

\section{POLYMER APPROACH TO THE QUANTUM MIXMASTER MODEL}
\label{QUANTUM}
We now analyze the quantum properties of the \textit{polymer} Mixmaster model. As in Sec.\ref{Qua}, one searches a solution for the wave function of the form \begin{equation}
\Psi(p_{\pm},\alpha) = \chi(\alpha)\psi(\alpha, p_{\pm}).
\end{equation}
In this case, one can choose to describe the $\chi(\alpha)$ component of the wave function in the $q$-polarization and the $\psi(\alpha, p_{\pm})$ component of the wave function in the $p$-polarization.
As in the semiclassical model, we choose to discretized the anisotropies ($\beta_{+},\beta_{-}$) leaving unchanged the characterization of the isotropic variable $\alpha$. 
Therefore, as in Sec.\ref{PQM}, one applies the formal substitution $\widehat{p}_{\pm}^{2}\rightarrow \frac{2}{a^{2}}\left[1-\cos (a p_{\pm})\right]$. Of course, the conjugated momenta $p_{\alpha}$ have a well-defined operator of the form $\widehat{p}_{\alpha} = -i\partial_{\alpha}$. This way, we can obtain the WDW equation for the \textit{polymer} Mixmaster model writing the quantum version of superHamiltonian in (\ref{vinSHpoly}), that is
\begin{multline}
\label{WDWpoly}
[ -\partial ^{2}_{\alpha}+\frac{2}{a^{2}}\left(1-\cos (a p_{+})\right)+\frac{2}{a^{2}}\left(1-\cos (a p_{-})\right) +\\ +\frac{3(4\pi)^{4}}{k^{2}}e^{4\alpha}V(\beta _{\pm}) ]\Psi(p_{\pm},\alpha) = 0.
\end{multline}
The conservation of quantum numbers associated to the anisotropies, as obtained by C.W.Misner in the standard quantum theory (see Eq.(\ref{numerioccupazione})), is essentially based on a fundamental propriety of the Mixmaster Model: the presence of chaos. Nevertheless, as in Sec.\ref{SEMI}, the chaos is removed for discretized anisotropies of the Universe. This way, one cannot obtain for the modified theory a conservation law towards the singularity as in the standard case. 
For a quantum description, the \textit{polymer} wavepackets for the theory are needed. By a semiclassical analysis of the relational motion between the wall and the particle, as in Sec.\ref{SEMI}, the \textit{polymer} modification implies for the particle different condition for the reach of the potential wall. This way, it behaves as a free particle (no potential case $V=0$) or as a particle in a box (infinitely steep potential well case). In this Section, we make use of the the adiabatic approximation (\ref{adiabatic}) as in the standard case. Following the same procedure of Sec.\ref{Qua}, the \textit{polymer} WDW equation reduces to an eigenvalue problem associated to the ADM Hamiltonian.
\subsection{The free motion}
\label{Free}
In the free particle case, the potential term $V(\beta_{\pm})$ is negligible in the WDW equation. As in Sec.\ref{Qua}, condition (\ref{adiabatic}) is applied  to Eq.(\ref{WDWpoly}), and the following free-particle eigenvalue problem is obtained
\begin{multline}
\label{quadratopoly}
 \widehat{H}^{2}_{poly}\psi(p_{\pm}) = k^{2}\psi(p_{\pm}) = \\ =\left[ \frac{2}{a^{2}}(2 - \cos(ap_{+}) - \cos(ap_{-})) \right]\psi(p_{\pm}).
\end{multline}
From the structure of the eigenvalue problem (\ref{quadratopoly}), one can write $\widehat{H^{2}}_{poly}=\widehat{H^{2}}_{+}+\widehat{H^{2}}_{-}$. As a consequence, it is possible to describe the anisotropic wave function as $\psi(p_{\pm}) = \psi_{+}(p_{+})\psi_{-}(p_{-})$. This way, one obtains the two independent eigenvalue problems
\begin{equation}
\begin{split}
\label{EFPR}
&(\widehat{H_{+}^{2}} - k_{+}^{2})\psi_{+}(p) =\left[\frac{2}{a^{2}}[1 - \cos(ap_{+})] - k_{+}^{2}\right]\psi_{+}(p) = 0, \\
&(\widehat{H_{-}^{2}} - k_{-}^{2})\psi_{-}(p) =\left[\frac{2}{a^{2}}[1 - \cos(ap_{-})] - k_{-}^{2}\right]\psi_{-}(p) = 0.
\end{split}
\end{equation}
where $k^{2} = k_{+}^{2} + k_{-}^{2}$. These eigenvalue problems can be treated as in Sec.\ref{FPP} and, by a similar procedure, one can easily verify that the momentum wave functions $\psi_{+}(p)$ and $\psi_{-}(p)$ have the form
\begin{equation}
\begin{split}
\label{beta+}
&\psi_{+}(p_{+}) = A\delta(p_{+} - p_{a}^{+}) + B\delta(p_{+} + p_{a}^{+}), \\ &\psi_{-}(p_{-}) = C\delta(p_{-} - p_{a}^{-}) + D\delta(p_{-} + p_{a}^{-}),
\end{split}
\end{equation}
where $A,B,C,D$ are integration constants and $p_{a}^{+}$,$p_{a}^{-}$ are defined as
\begin{equation}
\begin{split}
\label{reldis+}
&p_{a}^{+} = \frac{1}{a} \arccos\left(1 - \frac{k_{+}^{2}a^{2}}{2}\right), \\ &p_{a}^{-} = \frac{1}{a} \arccos\left(1 - \frac{k_{-}^{2}a^{2}}{2}\right).
\end{split}
\end{equation}
From Eq.'s (\ref{EFPR}), the eigenvalue $k^{2}$ is given by
\begin{multline}
k^{2} = k_{+}^{2}+k_{-}^{2} = \\ = \frac{2}{a^{2}} \left[ 2 - \cos(ap_{+}) - \cos(ap_{-})\right] \leq k^{2}_{max} =  \frac{8}{a^{2}},
\end{multline}
i.e. a bounded and continous eigenvalue is found.\\
Now one can obtain $\psi(\beta_{\pm})$ by performing a Fourier trasform for $\psi(p_{\pm}) = \psi_{+}(p_{+})\psi_{-}(p_{-})$, such that
\begin{multline}
\label{autolibera}
\psi_{k}(\beta_{\pm}) = \int\int dp_{+}dp_{-} \psi(p_{\pm}) e^{ip_{+}\beta_{+}}e^{i p_{-}\beta_{-}} = \\ =  C_{1} e^{ip_{a}^{+}\beta_{+}}e^{ip_{a}^{-}\beta_{-}} + C_{2} e^{ip_{a}^{+}\beta_{+}}e^{-ip_{a}^{-}\beta_{-}} + \\ +C_{3} e^{-ip_{a}^{+}\beta_{+}}e^{ip_{a}^{-}\beta_{-}} + C_{4} e^{-ip_{a}^{+}\beta_{+}}e^{-ip_{a}^{-}\beta_{-}},
\end{multline}
where $C_{1}=AC$,  $C_{2}=AD$, $C_{3}=BC$, $C_{4}=BD$. 
We are now able to build up the \textit{polymer} wave packet for the wave function of the Universe. We choose to integrate the packet on the energies $k_{+},k_{-}$. As a consequence of the modified dispersion relations (\ref{reldis+}), the energies eigenvalues $k_{+},k_{-}$ can only take values within the interval $[-\frac{2}{a},+\frac{2}{a}]$. Therefore, we have 
\begin{equation}
\label{paccfree}
\Psi(\beta _{\pm},\alpha) = \iint_{-\frac{2}{a}}^{\frac{2}{a}} dk_{\pm}A(k_{\pm})\psi _{k_{\pm}}(\beta_{\pm})\chi(\alpha),
\end{equation}
where $A(k_{+},k_{-}) = e^{-\frac{(k_{+}-k_{+}^{0})^{2}}{2 \sigma _{+} ^{2}}}e^{-\frac{(k_{-}-k_{-}^{0})^{2}}{2 \sigma _{-} ^{2}}}$ is a Gaussian weighting function, $\sigma^{2}_{\pm}$ are the variances along the two directions ($\beta_{+}$,$\beta_{-}$) and $k_{\pm}^{0}$ are the energies eigenvalues around which we build up the wave packet. Let us note from Eq.(\ref{paccfree}) that the polymer structure modifies the standard wave packet related to the plane wave in terms of the anisotropies component as a consequence of Eq.'s(\ref{reldis+}), i.e. the modified dispersion relations.
\\
The shape for the isotropic component of the wave function in the free particle case is $\chi(\alpha)= e^{-i\int_{0}^{\alpha}kdt} = e^{-i\sqrt{k_{+}^{2} + k_{-}^{2}}\alpha} $. This shape is a solution of the WDW equation $\partial^{2}\chi(\alpha) + k^{2}\chi(\alpha) = 0$ obtained by the application of the adiabatic approximation (\ref{adiabatic}). Furthermore, the self-consistence of this approximation is ensured.
\begin{figure*}[tb]
\includegraphics[scale=.45]{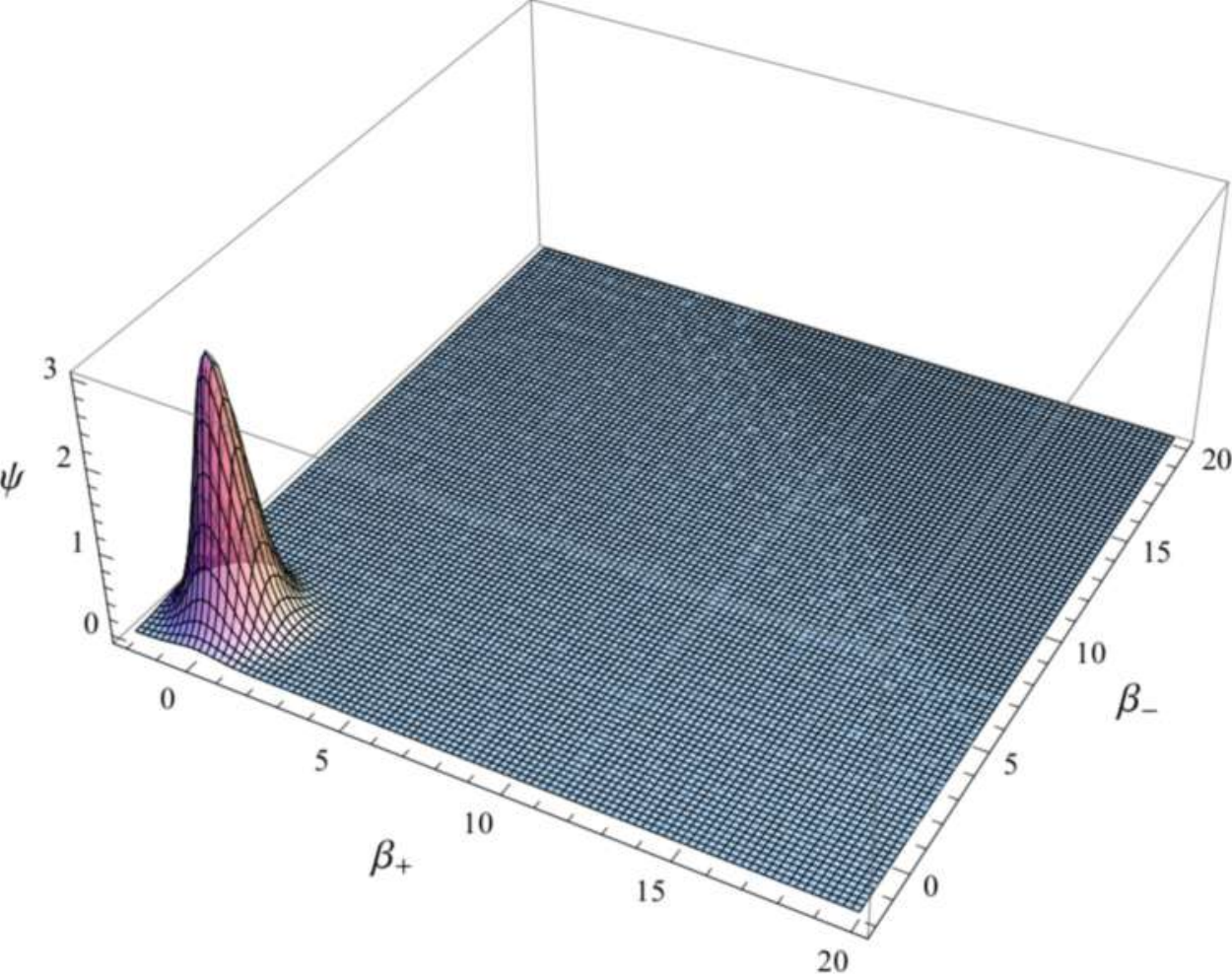}
\includegraphics[scale=.45]{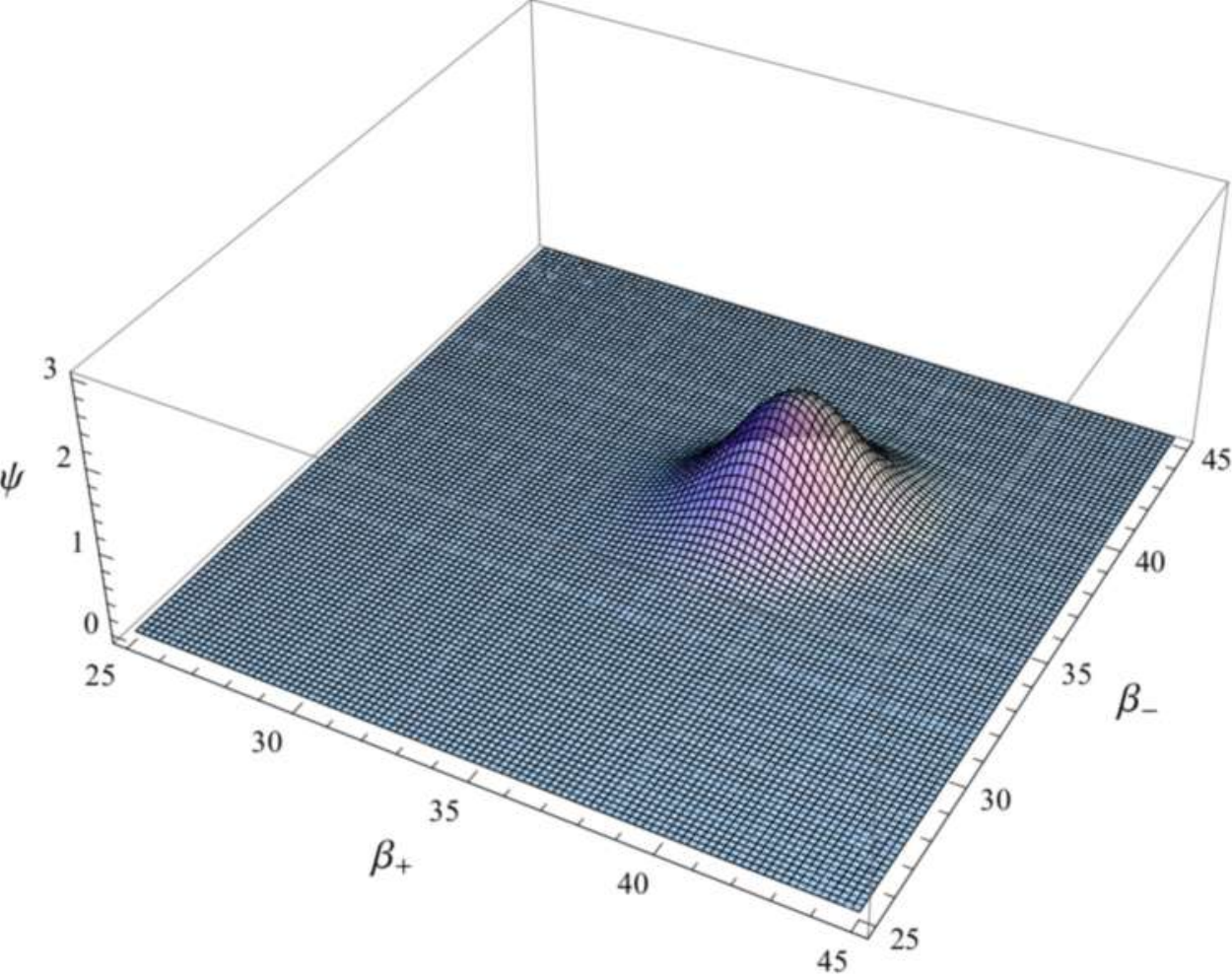}
\includegraphics[scale=.45]{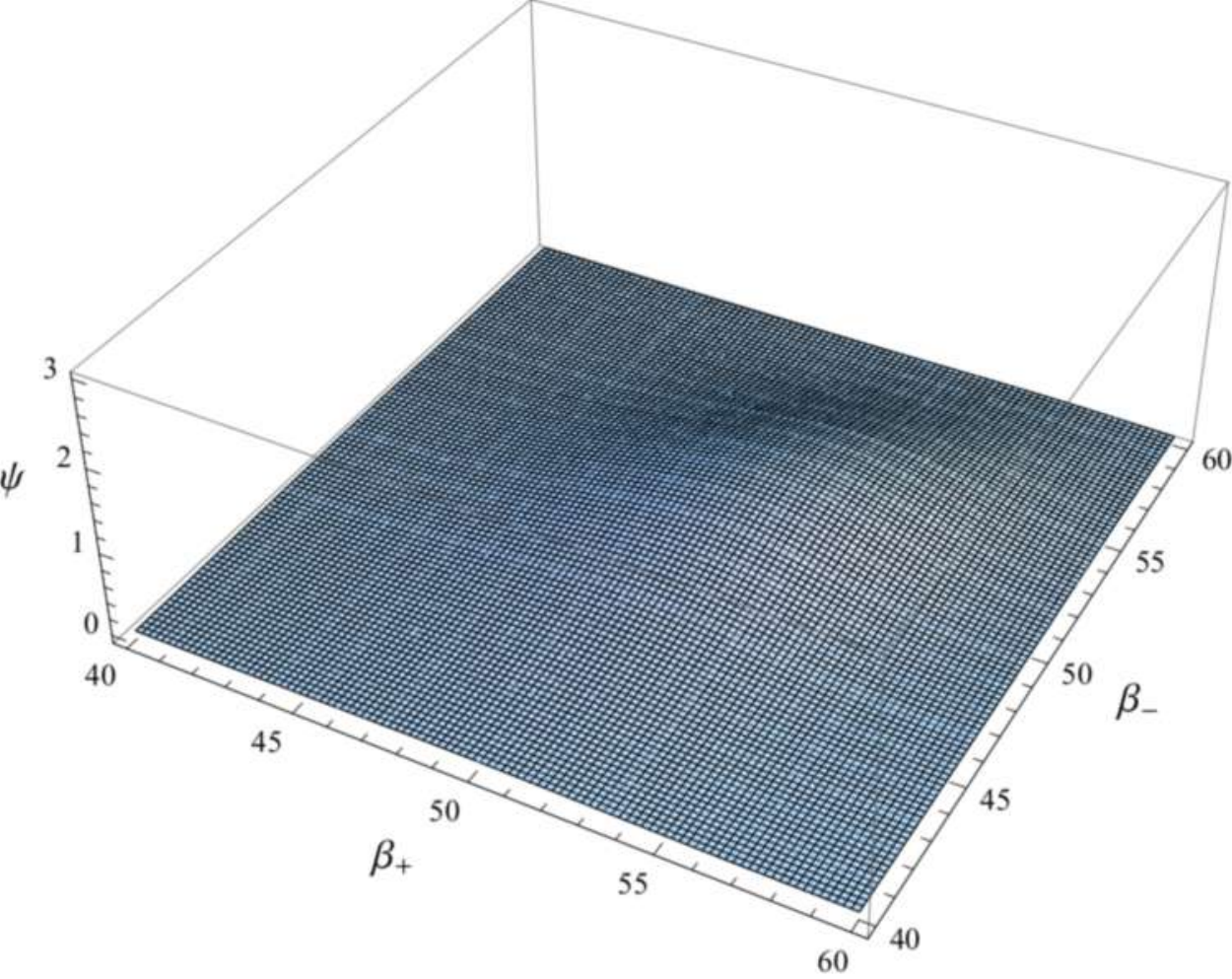}
\includegraphics[scale=.55]{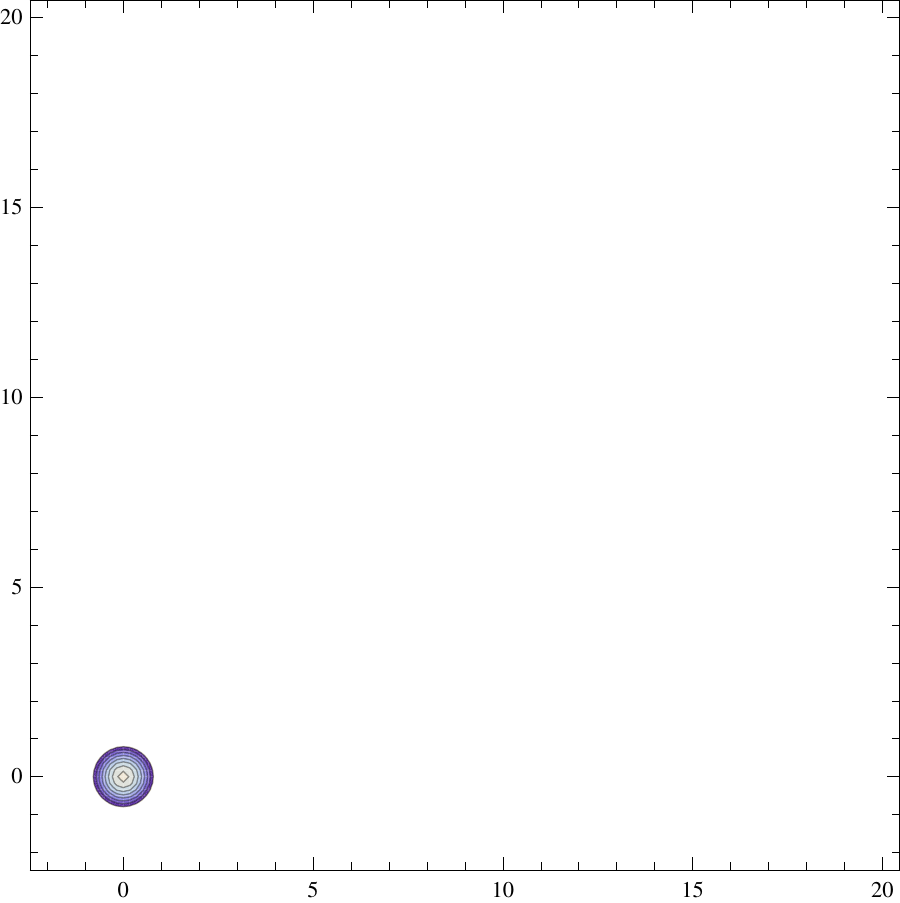}\quad \quad
\includegraphics[scale=.55]{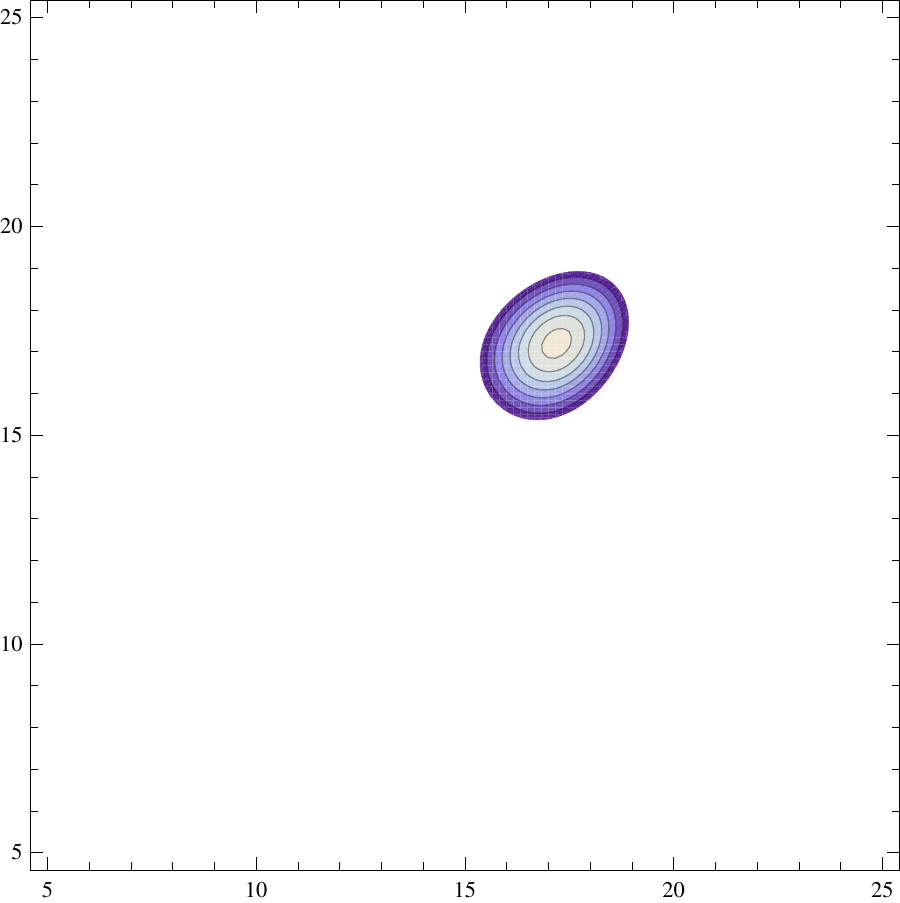}\quad \quad
\includegraphics[scale=.55]{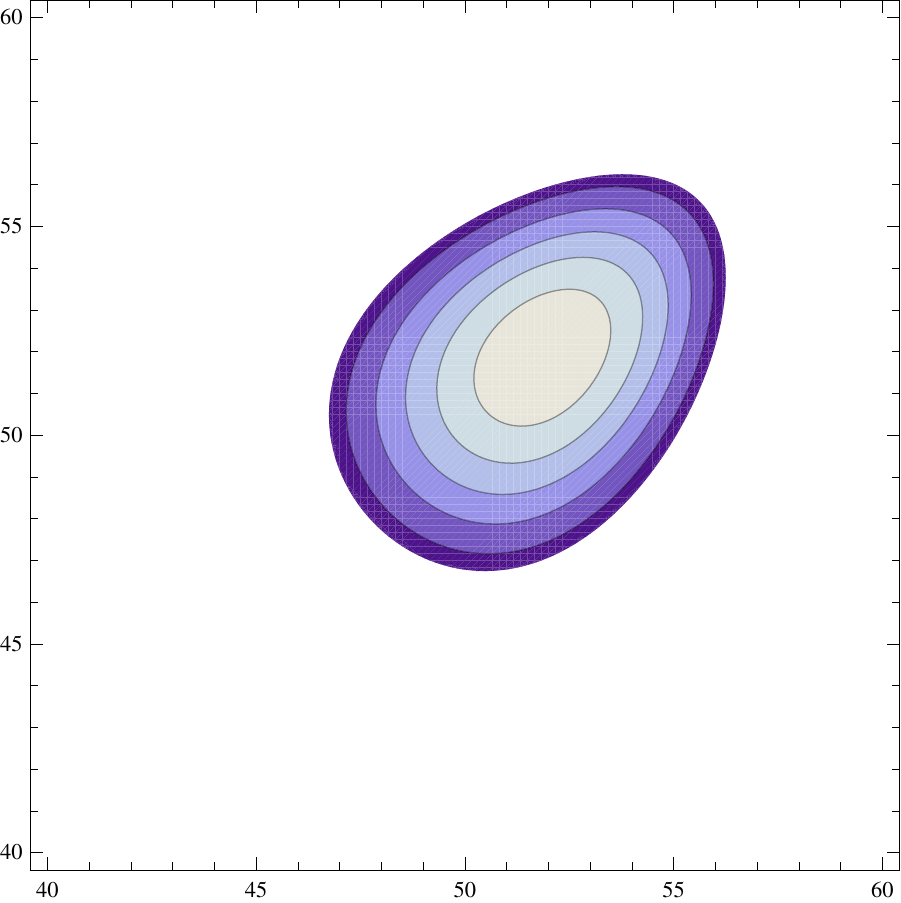}
\caption{\footnotesize "(Color online)". The evolution of the \textit{polymer} wave packet $|\Psi(\alpha,\beta_{\pm})|$(upper row) and its full width at half maximum (lower row) for the free particle case respectively for the values of $|\alpha | =0,50,150$. The numerical integration is done for this choice of parameters: $a=0.07,k_{+}=k_{-}=25, \sigma_{+}=\sigma_{-}=0.7$. They select an initial semiclassical condition of a particle with a velocity smaller than the wall velocity. It is worth noting that the  particular choice of the parameters couple ($a,\sigma_{\pm}$) is done because this way the condition $a<<\frac{1}{\sigma_{\pm}}$ is valid. It is referred to the condition that the typical \textit{polymer} scale $a$ be much smaller than the characteristic width of the wave packet $\frac{1}{\sigma_{\pm}}$. }
\label{fig:polylib}
\end{figure*}
\begin{figure*}[tb]
\centering
\includegraphics[scale=.67]{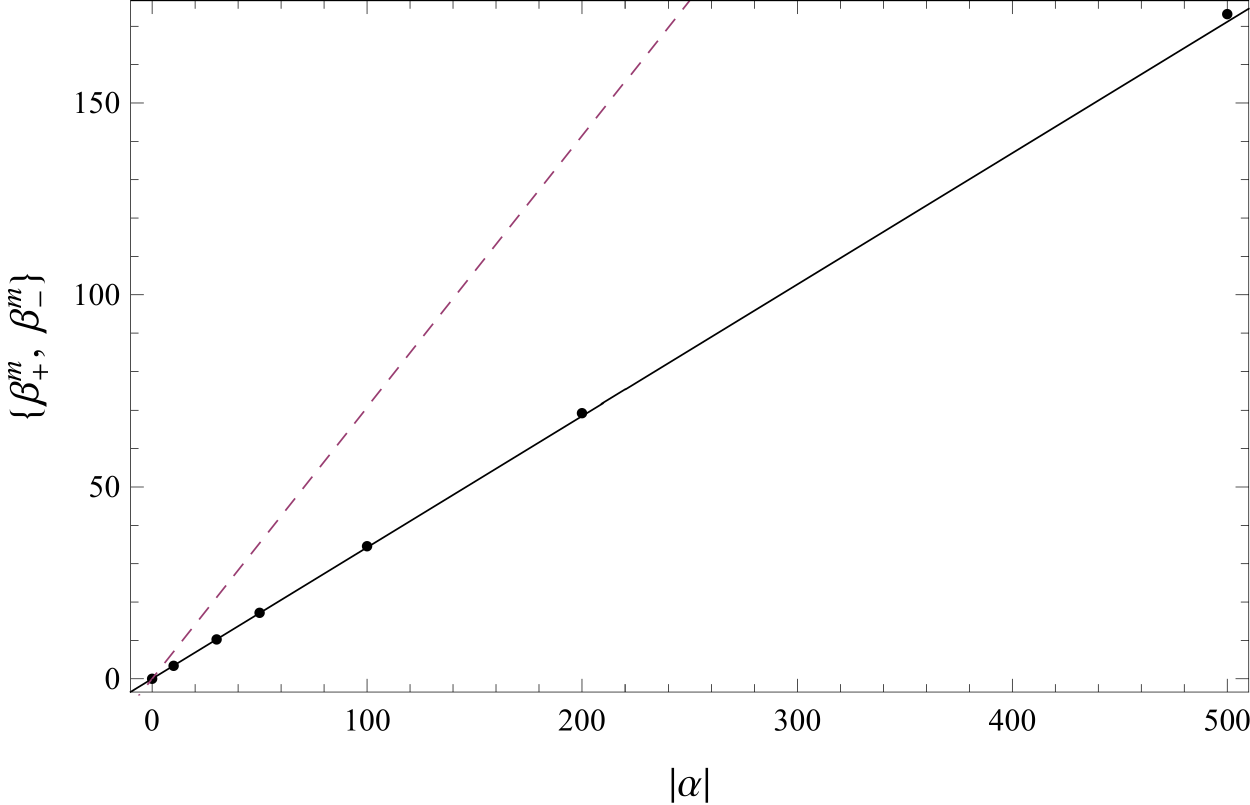}\qquad
\includegraphics[scale=.67]{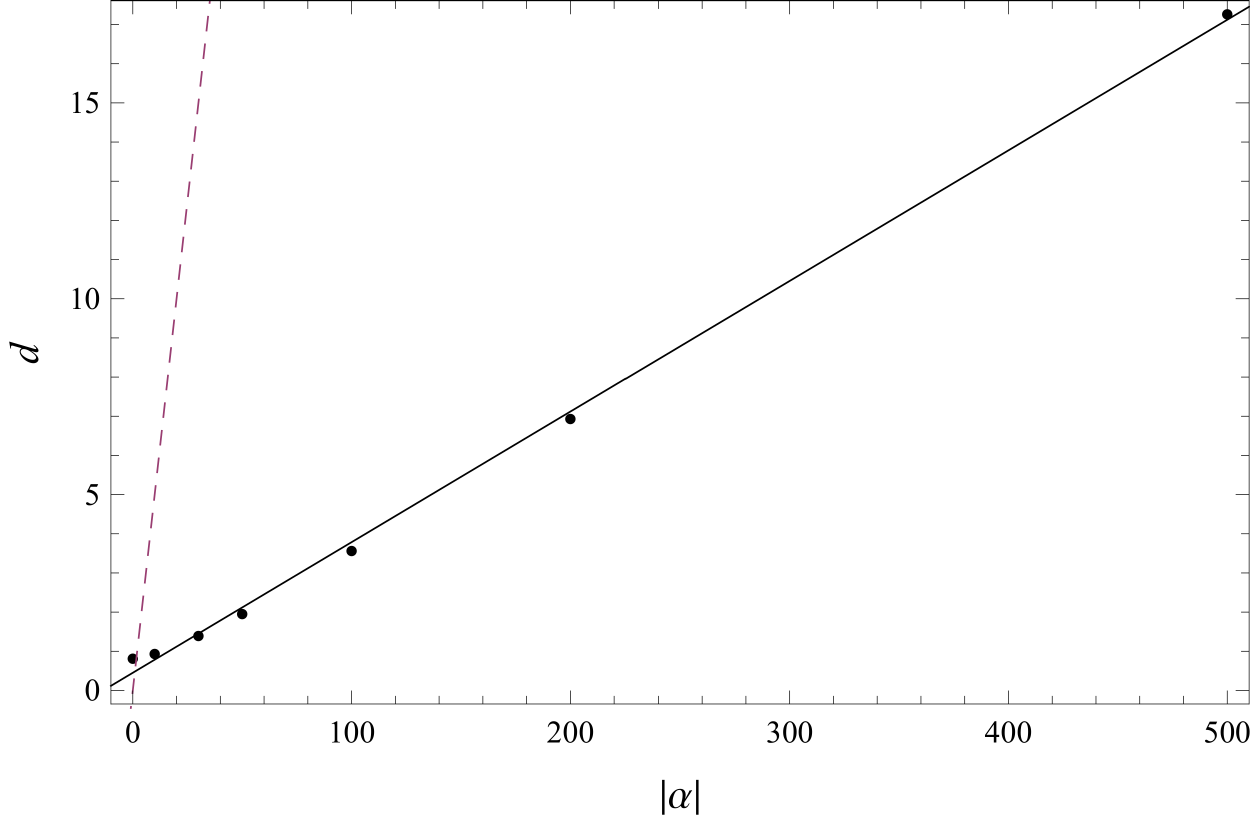}
\caption{\footnotesize "(Color online)". The solid line in the first graph represents the \textit{polymer} semiclassical trajectory identified by the choice of the initial conditions. The dashed line represents the classical trajectory followed by a wave packet build up in the same way of Sec.\ref{Free} but
starting from classical superHamiltonian constrain (\ref{vincolo misner}). The points in the second graph represent the evolution of the spread $d$ as a function of $|\alpha|$. The solid line represents the best fit for the points while the dashed line represents the evolution of the wall position $|\beta_{w}|=\frac{1}{2}|\alpha|$.}
\label{traspre}
\end{figure*}
\subsection{Particle in a box}
\label{Box}
We analyze the problem of a particle in a box according to the Misner hypothesys about the substitution of the triangular box by a square domain having the same area $L^{2}$, as in Sec.\ref{Qua}. Furthermore, following the semiclassical results in Sec.\ref{SEMI}, one takes into account the outside wall velocity defining the side of square box $L$ as
\begin{equation}
\label{lato}
L(\alpha) = L_{0} + | \alpha |,
\end{equation}
where $L_{0}$ is the side of the square box when $\alpha=0$.
Proceding in the same way as in Sec.\ref{PPB}, the potential has the well-known form 
\begin{equation}
\label{potbox}
V(\beta_{\pm}) = \begin{cases} \infty, & \beta_{\pm} > \frac{L(\alpha)}{2} \quad,\quad \beta_{\pm}<-\frac{L (\alpha)}{2}  \\ 0, & -\frac{L(\alpha)}{2} < \beta_{\pm} <\frac{L(\alpha)}{2}
 \end{cases}.
\end{equation}
We can obtain a solution for $\psi(\beta_{\pm})$ in the same way of Sec.\ref{Free}, recalling that the potential form (\ref{potbox}) implies this kind of boundary conditions for $\psi(\beta_{\pm})$ along the two directions
\begin{equation}
\label{uno}
\psi _{\pm} \left( -\frac{L_{0}}{2}-\frac{\alpha}{2} \right) = \psi _{\pm} \left( +\frac{L_{0}}{2}+\frac{\alpha}{2} \right) = 0.
\end{equation}
When one applies the conditions (\ref{uno}) separately along the two directions $(\beta_{+},\beta_{-})$, one obtains
\begin{equation}
\begin{split}
\label{sca+}
&\psi _{+}(\beta_{+})=A\left[e^{\frac{i n \pi \beta_{+}}{L_{0}+\alpha} } - e^{\frac{-i n \pi \beta_{+}}{L_{0}+\alpha} }e^{-i n \pi } \right], \\
&\psi_{-}(\beta_{-})=B\left[e^{\frac{i m \pi \beta_{-}}{L_{0}+\alpha} } - e^{\frac{-i m \pi \beta_{-}}{L_{0}+\alpha} }e^{-i m \pi } \right].
\end{split}
\end{equation}
This way, $\psi(\beta _{\pm})$ is the product of the two separate wave functions $\psi _{+}(\beta_{+})$ and $\psi _{-}(\beta_{-})$. Thus, one gets\footnote{It is possible to evaluate the costant $AB$ by requesting that $|\psi_{n,m}(\beta_{\pm})|^{2}=1$ over all the square box. This way, $AB = \frac{1}{2(L_{0}+\alpha)}$ is obtained.}
\begin{multline}
\psi_{n,m}(\beta _{\pm},\alpha) = \psi _{+}(\beta_{+})\psi _{-}(\beta_{-}) = \\ = \frac{1}{2(L_{0}+\alpha)} \left[e^{\frac{i n \pi \beta_{+}}{L_{0}+\alpha} } - e^{\frac{-i n \pi \beta_{+}}{L_{0}+\alpha} }e^{-i n \pi } \right] \times \\ \times \left[e^{\frac{i m \pi \beta_{-}}{L_{0}+\alpha} } - e^{\frac{-i m \pi \beta_{-}}{L_{0}+\alpha} }e^{-i m \pi } \right],
\end{multline}
where $A,B$ are integration constants and $(n,m) \in \mathbb{Z}$ are quantum numbers associated anisotropy degrees of freedom. Due to the presence of the integers quantum numbers $(n,m)$, a bounded and discrete eigenvalue spectrum
\begin{multline}
\label{autovalori}
k^{2} = k^{2}_{+} + k^{2}_{-}= \\ =\frac{2}{a^{2}}\left[ 2-\cos \left( \frac{a n \pi }{L_{0}+\alpha} \right) - \cos \left( \frac{a m \pi }{L_{0}+\alpha} \right) \right]
\end{multline}
is obtained.\\
As in the free particle case, one builds the \textit{polymer} wave packet. However, in this case, one cannot integrate on a limited domain of energies $k_{\pm}$, and a sum over all quantum numbers $n,m$ between $-\infty$ and $\infty$ i necessary. This way,
\begin{multline}
\label{paccbox}
\Psi(\beta _{\pm},\alpha) = \sum_{n,m=-\infty}^{+\infty}B(n,m)\psi _{n,m}(\beta _{\pm},\alpha) \times \\ \times e^{-i\begin{matrix}
\int_{0}^{\alpha}\sqrt{ \frac{2}{a^{2}}\left[2-\cos\left(\frac{an\pi}{L_{0}+t}\right)-\cos\left(\frac{am\pi}{L_{0}+t}\right) \right]}dt,
\end{matrix}}
\end{multline}
where $B(n,m)=e^{-\frac{(n-n^{*})^{2}}{2 \sigma _{+} ^{2}}}e^{-\frac{(m-m^{*})^{2}}{2 \sigma _{-} ^{2}}}$ is a Gaussian weighting function and $n^{*},m^{*}$ are the quantum numbers around which we build up the wave packet.\\
Let us note that, differently from the free particle case, the presence of the polymer structure modifies the standard wave packet related to a particle in a box in terms of the isotropic components. It happens because, in the wave packet (\ref{paccbox}), the energies $k_{\pm}$ are expressd through ($n,m$), namely the quantum numbers associated to the anisotropies.\\
As from Eq.(\ref{paccbox}), one chooses a shape for the isotropic component 
\begin{equation}
\label{iso}
\chi(\alpha)= e^{-i\int_{0}^{\alpha}k(t)dt} = e^{-i\int_{0}^{\alpha}\sqrt{ \frac{2}{a^{2}}\left[2-\cos\left(\frac{an\pi}{L_{0}+t}\right)-\cos\left(\frac{am\pi}{L_{0}+t}\right) \right]}dt}.
\end{equation}
In this case, Eq.(\ref{iso}) is a solution of the WDW equation $\partial^{2}\chi(\alpha) + k(\alpha)^{2}\chi(\alpha) = 0$ obtained by means of the adiabatic approximation (\ref{adiabatic}) in the asymptotic limit $\alpha\rightarrow-\infty$. 
In this limit, the self-consistence of the adiabatic approximation is ensured. The form of the isotropic component of the wave function (\ref{iso}) is also an exact solution for the Schr\"odinger equation associated to the ADM reduction.
\begin{figure*}[tb]
\includegraphics[scale=.45]{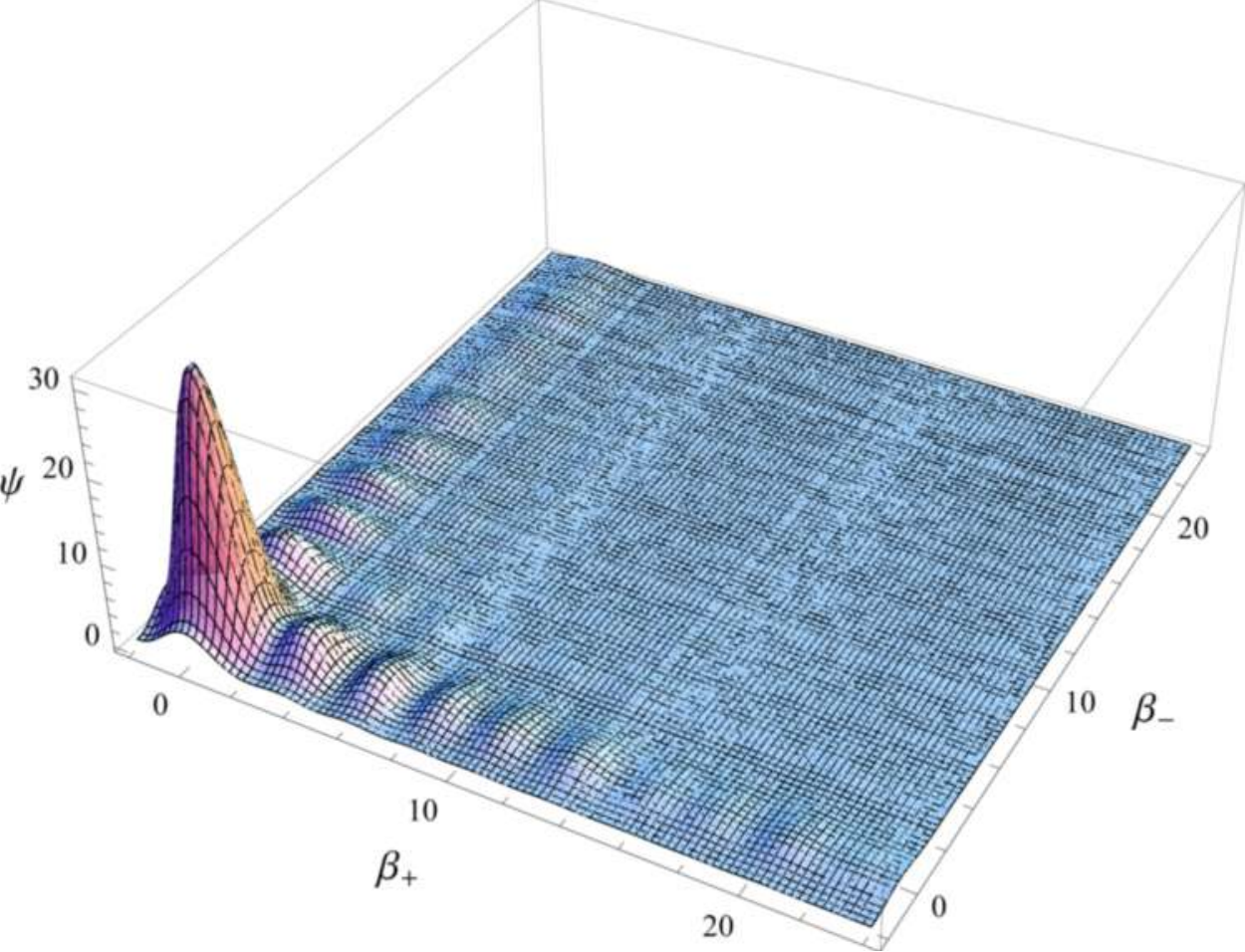}
\includegraphics[scale=.45]{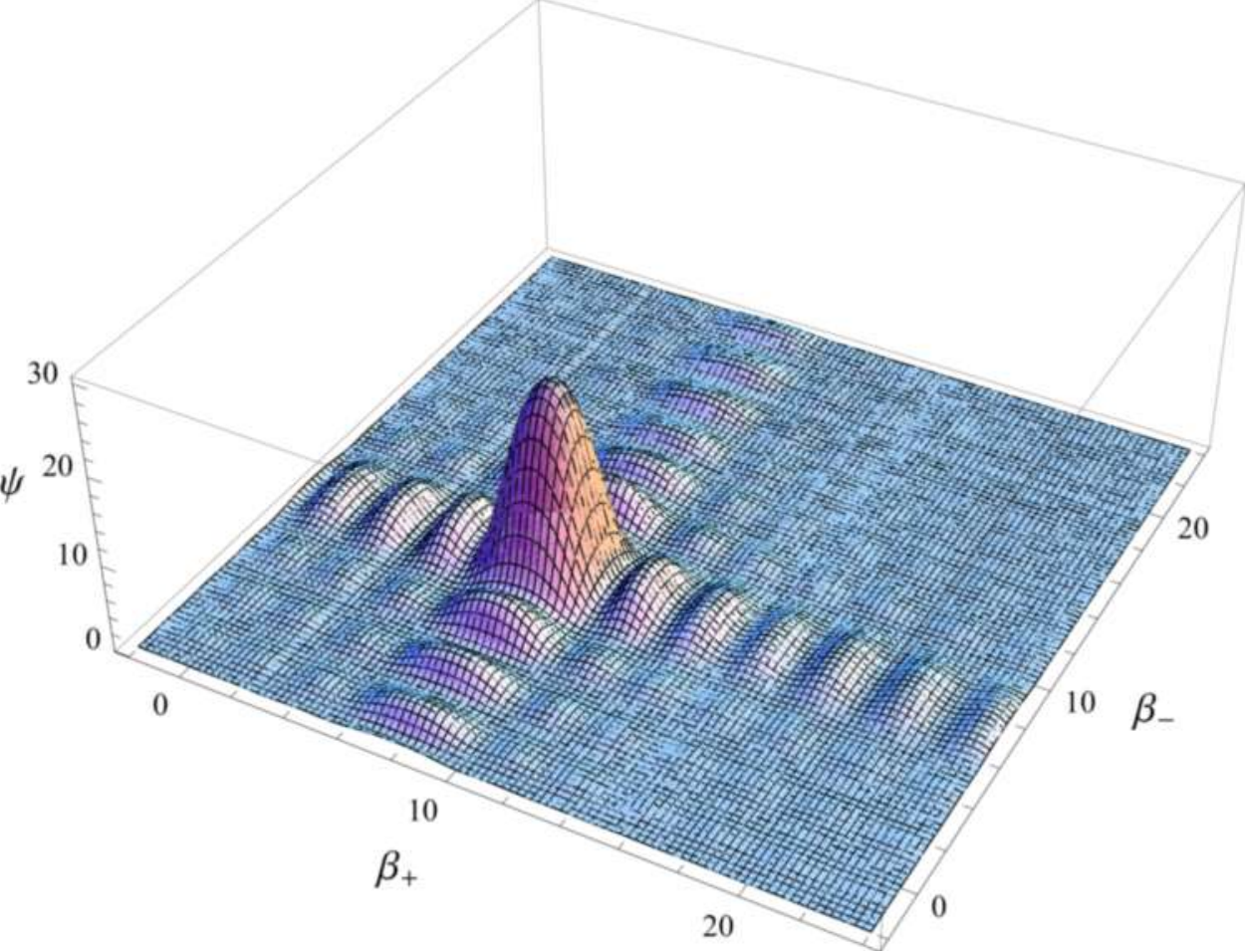}
\includegraphics[scale=.45]{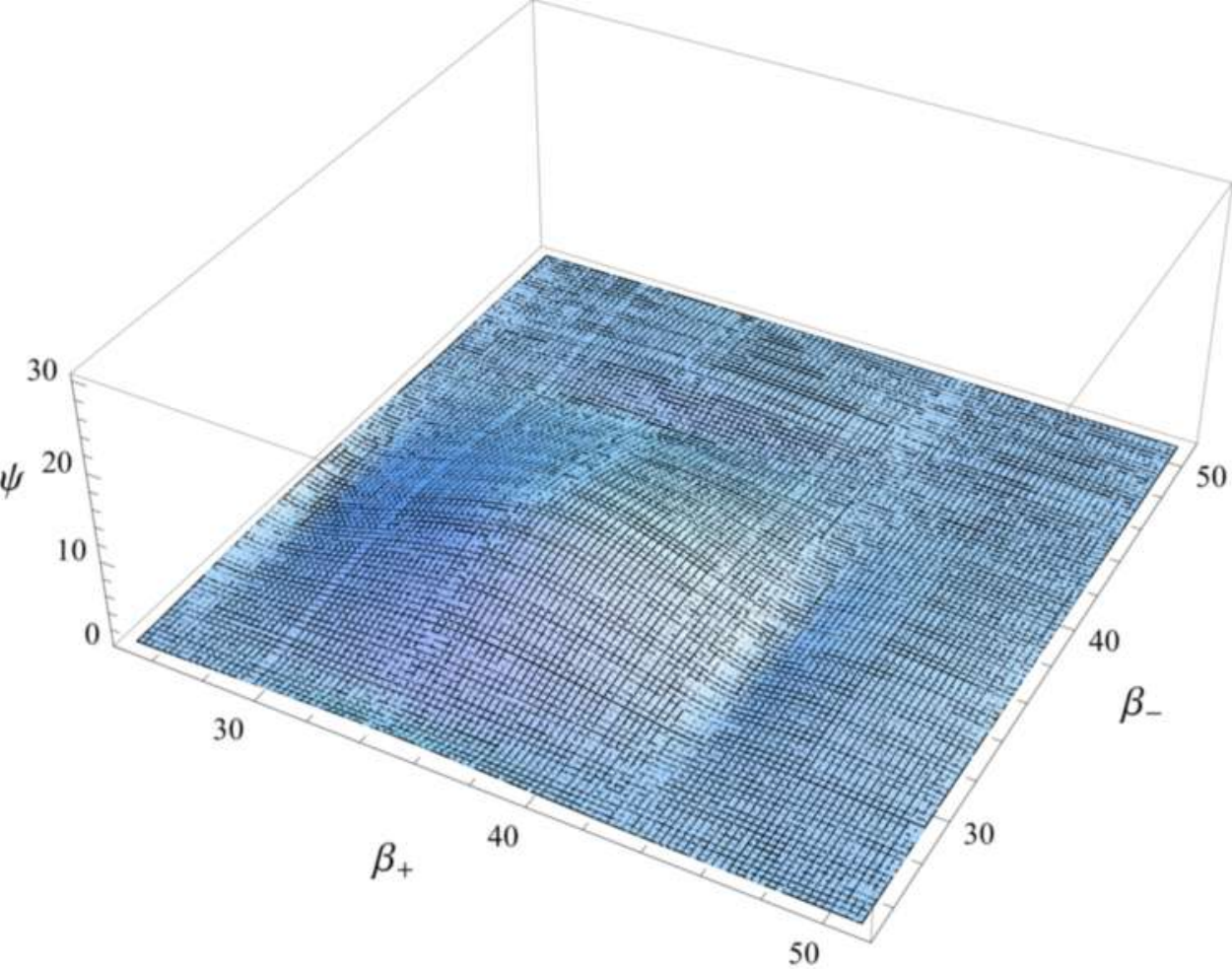}
\includegraphics[scale=.45]{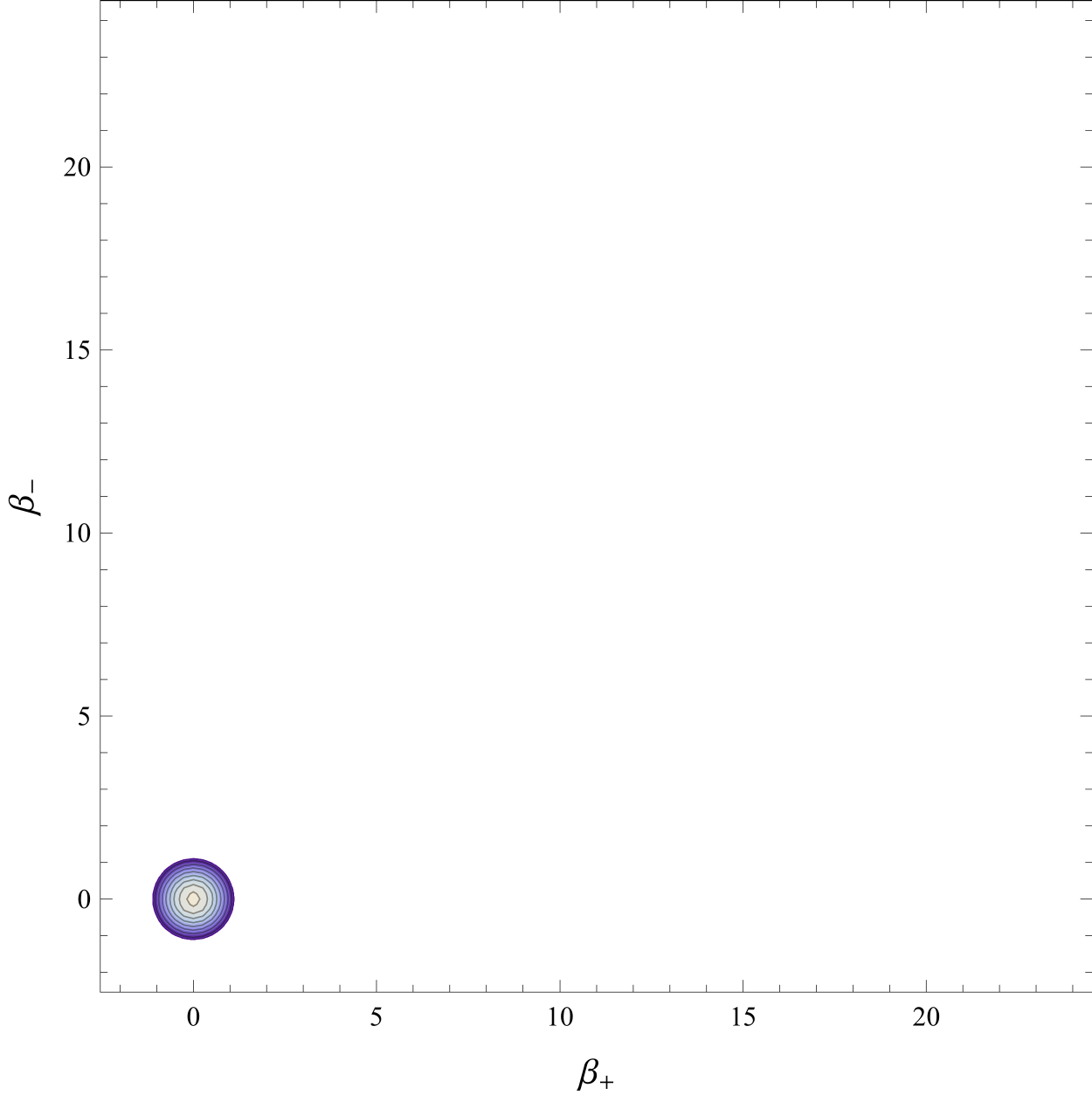}
\includegraphics[scale=.45]{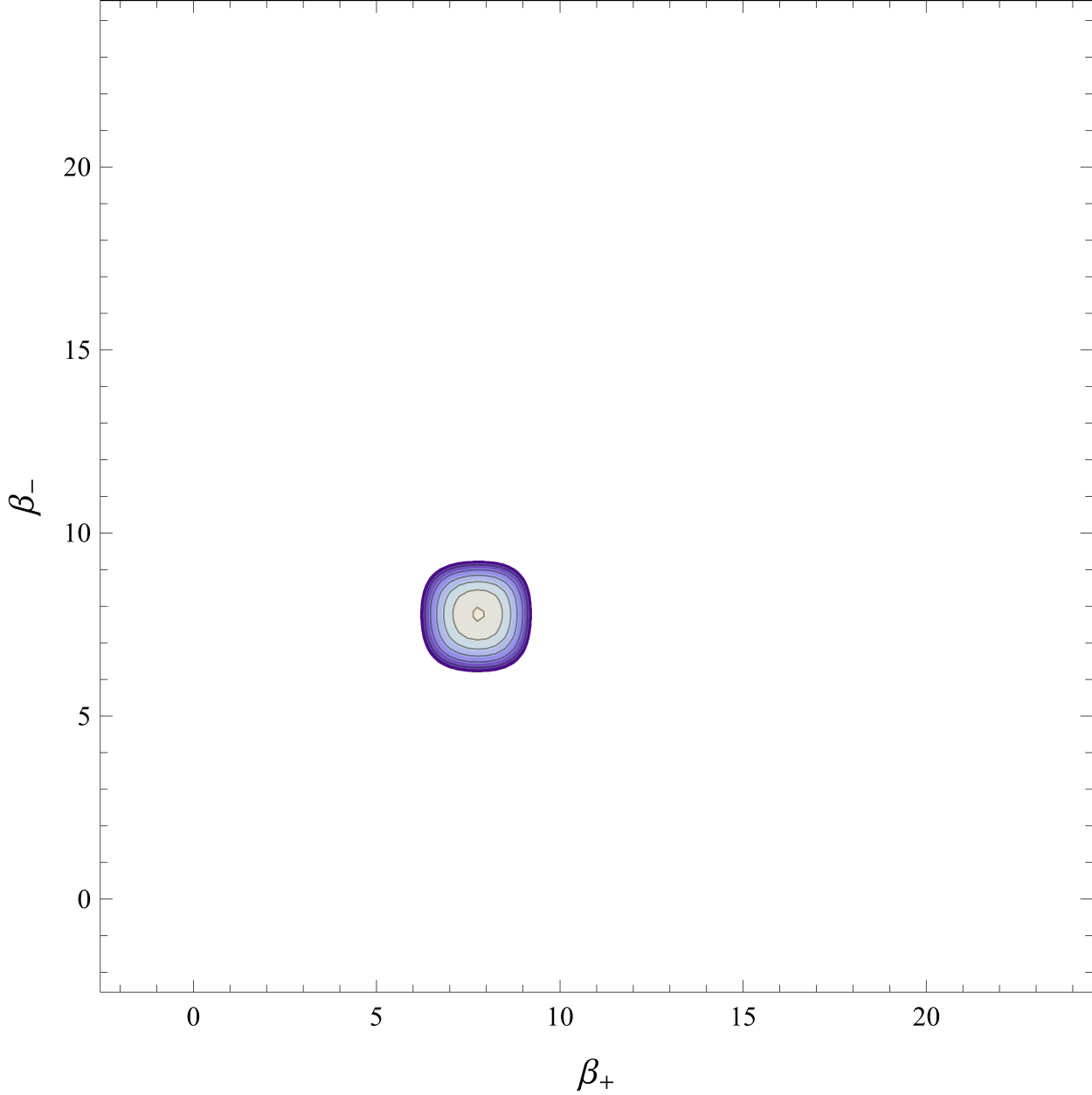} 
\includegraphics[scale=.45]{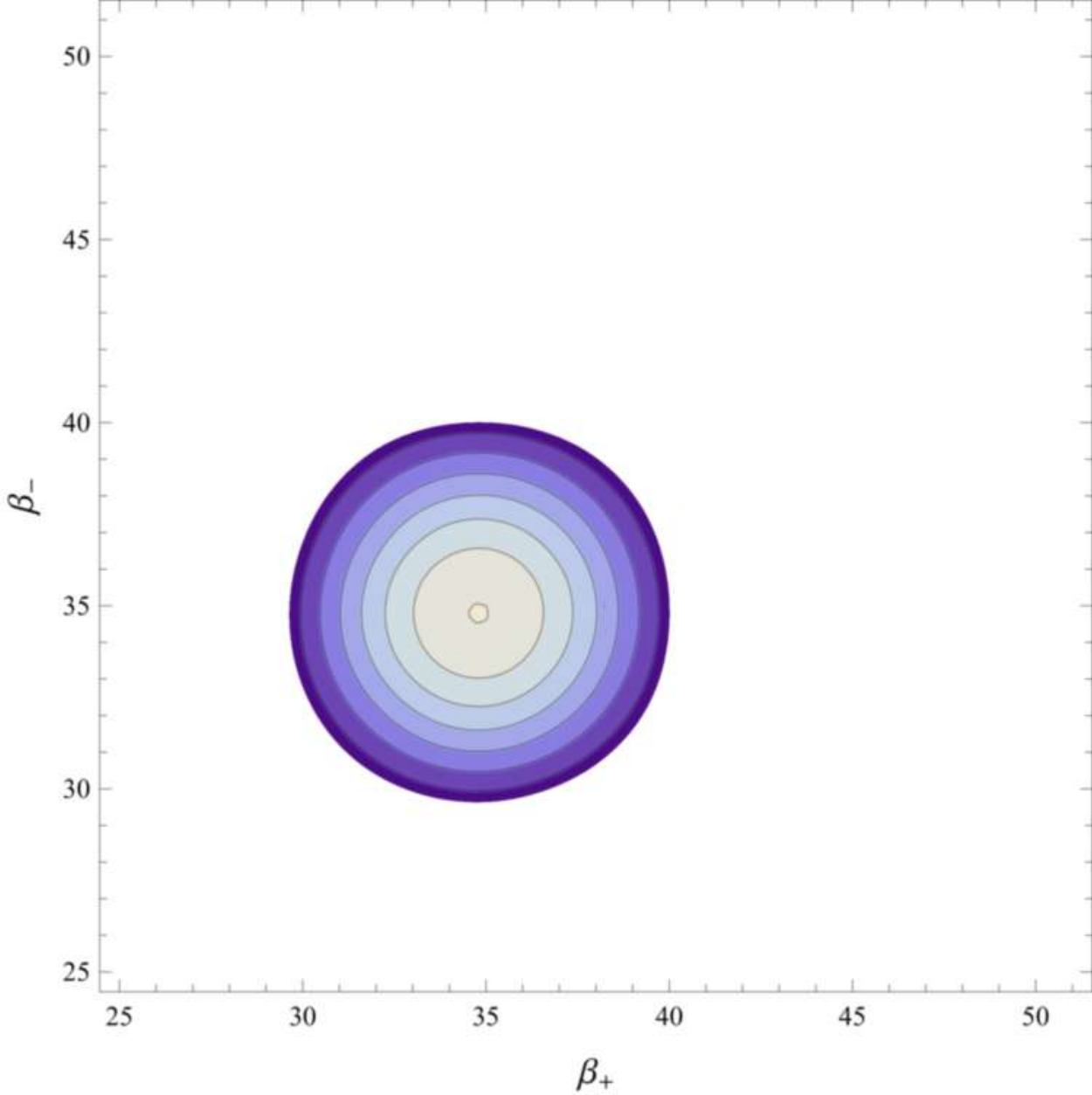}
\caption{\footnotesize "(Color online)". The evolution of the \textit{polymer} wave packet $|\Psi(\alpha,\beta_{\pm})|$(the first row) and its full width at half maximum (the second row) for the particle in a box case respectively for  $|\alpha | =0,20,200$. The numerical integration is done for this choice of parameters: $a=0.014,n^{*}=m^{*}=3000, \sigma_{+}=\sigma_{-}=50, L_{0}=52$. They select an initial condition of a particle inside a square box with velocity smaller than the wall velocity. This time, the particular choice of the parameters ($a,\sigma_{\pm},L_{0}$) it is done because this way the condition $a<<\frac{L(\alpha)}{\sigma_{\pm}}$ is valid. It concerns the condition that the typical \textit{polymer} scale $a$ is very smaller than  $\frac{L(\alpha)}{\sigma_{\pm}}$, i.e. the correct dimensional quantity related with the width of the wave packet. }
\label{fig:polybox}
\end{figure*}
\section{NUMERICAL ANALYSIS OF POLYMER WAVE PACKETS}
\label{NUM}
We dedicate this section to the discussion of the polymer wave packet for the Mixmaster towards the cosmological singularity. Both in the case of a free particle (\ref{paccfree}) and in the one of a particle in a box (\ref{paccbox}), it is not possible to perform an analytic integration for the wave packets. This way, in order to obtain the quantum behavior of the wave packets near the cosmological singularity, we evaluate them via numerical integrations.
\subsection{behavior of the free particle}
\label{FreePacket}
In the case of a free particle, we perform the numerical integration choosing the parameters which select semiclassical initial conditions concerning a particle with velocity smaller than the wall one ($r<\frac{1}{2}$).\\ One appreciates, in the first row of the Fig.(\ref{fig:polylib}), the behavior towards the singularity (formally for $|\alpha|\rightarrow\infty$) of the absolute value of the wave packet $|\Psi(\alpha,\beta_{\pm})|$ in Eq.(\ref{paccfree}) while, in the second row, the behavior towards the singularity of the full width at half maximum width.
It is interesting to study the evolution of $\beta_{\pm}^{m}$, i.e. the wave packet maximum position. This way, we can see which trajectory the wave packet follows towards the singularity. 
As we can see in the first graph in Fig.(\ref{traspre}), the behavior of the maximum position is completely overlapping the semiclassical trajectory selected by our choice of the initial conditions.
In this sense, the \textit{polymer} wave packet follows the semiclassical trajectory until the singularity. This feature is not undermined by the spread $d$ of the wave packet, i.e. the delocalization of the wave packet, as expressed by the distance between the maximum position of the wave packet and the edge of the region identified by the full width at half maximum. Obviously, one expects that the spread velocity is really smaller than the wall velocity. Otherwise, it would be possible for that the wave packet to reach the potential wall. In that case, the description of the quantum system with the wave packets for the free particle would not be correct. The second graph in Fig.(\ref{traspre}) represents the spread evolution, and we can see it follows a linear behavior (solid line) with a slope much smaller than $|\beta'_{w}|=\frac{1}{2}$, i.e. the one related to the behavior of the wall position (dashed line). This assures that the quantum representation of the system near the singularity for the free particle case is well described by the wave packet representation.
\subsection{behavior of the Particle in a box}
\label{BoxPacket}
\begin{figure}[tb]
\centering
\includegraphics[scale=.9]{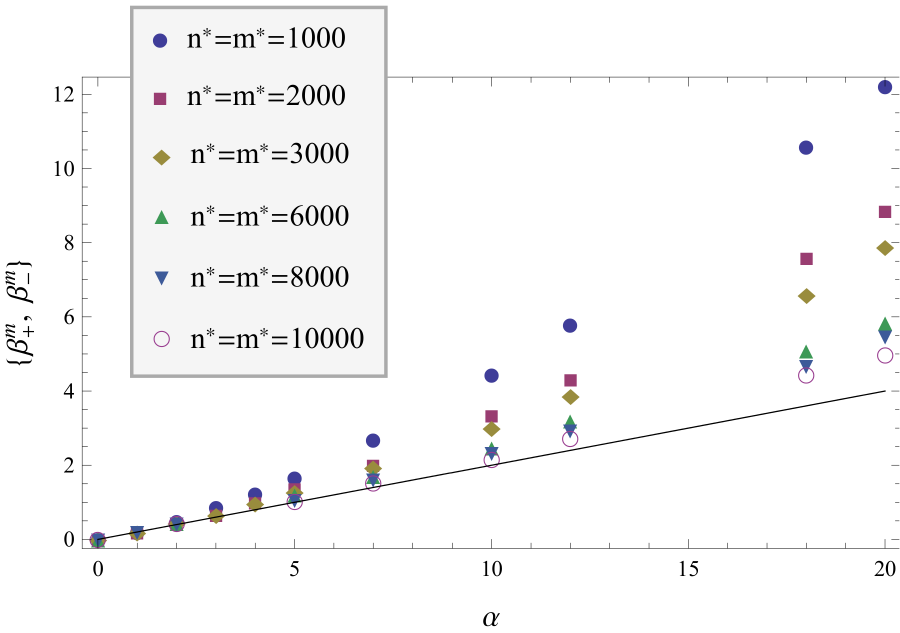}\quad
\includegraphics[scale=.63]{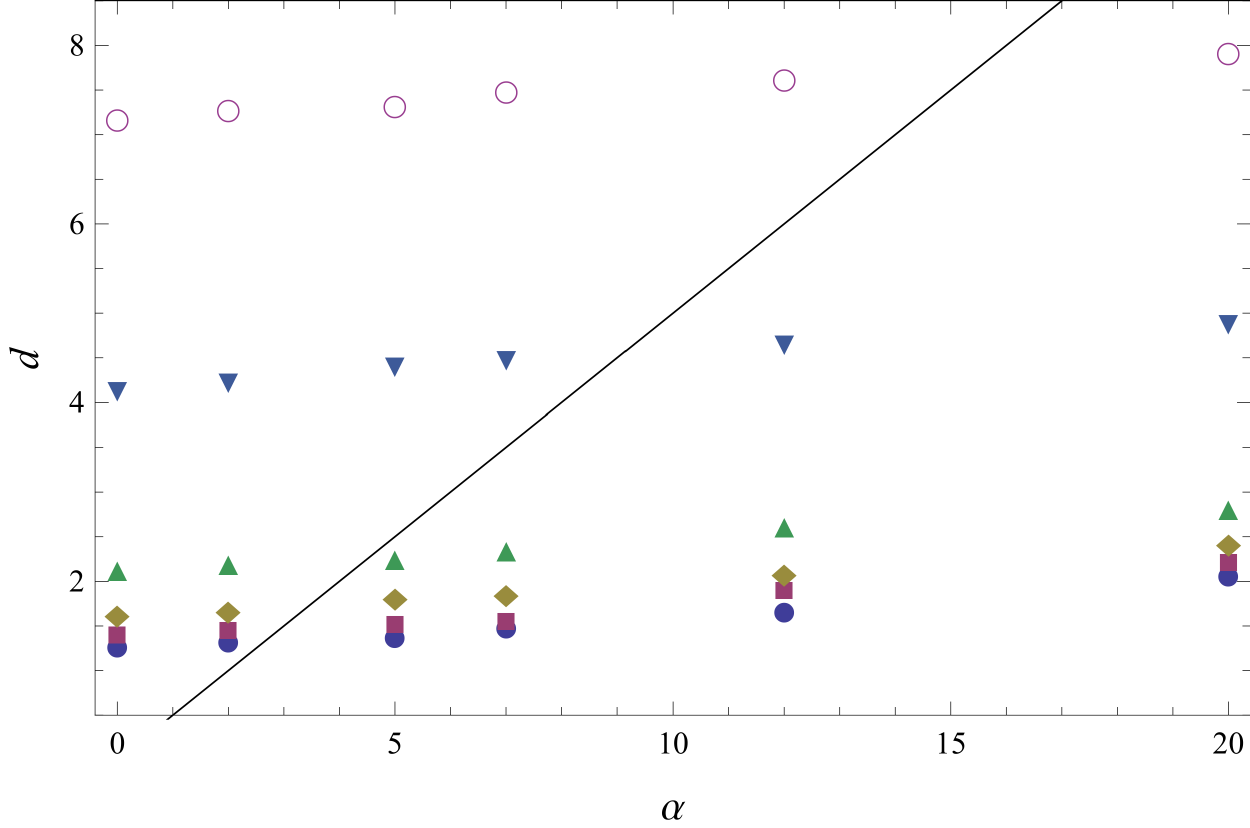}
\caption{\footnotesize "(Color online)". The points in the first graph represent the evolution of the wave packet maximum position $\beta_{\pm}^{m}$ as a function of $|\alpha|$ for all data sets. The solid line represents the \textit{polymer} semiclassical trajectory identified by the choice of the initial conditions. The points in the second graph represent the evolution of the spread $d$ as a function of $|\alpha|$ for all data sets. The solid line represents the evolution of the wall position $|\beta_{w}|=\frac{1}{2}|\alpha|$. As in the free particle case, the spread evolution follows a linear trend for all data sets and the slopes are really smaller than the one related to the trend of the wall position.}
\label{fig:picchi}
\end{figure}

\begin{figure}[tb]
\centering
\includegraphics[scale=.65]{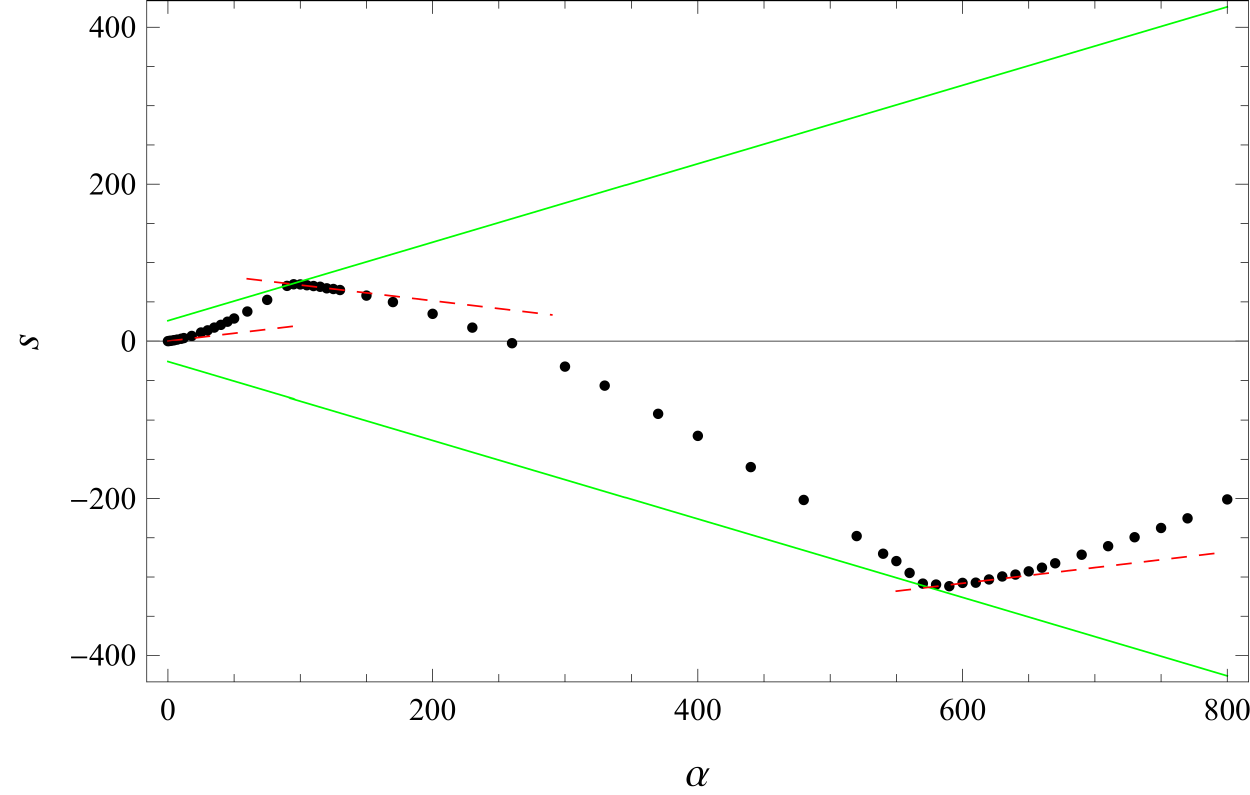}
\caption{\footnotesize "(Color online)". The points represent the evolution of the wave packet maximum position $\beta_{\pm}^{m}$ as a function of $|\alpha|$ for $a=00.14,n^{*}=m^{*}=3000, \sigma_{+}=\sigma_{-}=50, L_{0}=32$. The two solid lines represent the $\alpha$-evolution of the position of two opposite wall of the square box. At last, the dashed lines represent the \textit{polymer} semiclassical trajectory identified by the choice of the initial conditions that the wavepacket follow after each bounce for a finite $\alpha$-time.}
\label{fig:bounce}
\end{figure}

The numerical integration related to the \textit{polymer} wave packet (\ref{paccbox}) has to face a significant technical difficulty. As a consequence of Eq.(\ref{autovalori}), the conjugated momenta $p_{\pm}$ turn into a discretized variables. Therefore, we select for the particle in a box the initial semiclassical condition considering the substitution 
\begin{equation}
\label{disccon}
ap_{+} \rightarrow \frac{a n \pi}{L_{0}+\alpha} \quad , \quad  ap_{-} \rightarrow \frac{a m \pi}{L_{0}+\alpha}.
\end{equation}
It is worth noting that the initial condition of the particle depends on $\alpha$, such that one deals with a \textit{time}-dependent condition.
In this subsection, the influence of quantum numbers $n,m$ on the dynamics is investigated. For this reason, one introduces six data sets with different values of quantum numbers ($n^{*},m^{*}$) and box side $L_{0}$
\begin{equation}
\begin{split}
\label{casi}
&\begin{cases}
a = 0.014 \\ n_{0} = 1000 \\ m_{0} = 1000 \\ L_{0} = 17 \\ \sigma _{+} = 50 \\ \sigma _{-} = 50 
\end{cases}  \begin{cases}
a = 0.014 \\ n_{1} = 2000 \\ m_{1} = 2000 \\ L_{1} = 34 \\ \sigma _{+} = 50 \\ \sigma _{-} = 50 
\end{cases}  \begin{cases}
a = 0.014 \\ n_{3} = 3000 \\ m_{3} = 3000 \\ L_{3} = 52 \\ \sigma _{+} = 50 \\ \sigma _{-} = 50 
\end{cases} \\ &\begin{cases}
a = 0.014 \\ n_{4} = 6000 \\ m_{4} = 6000 \\ L_{4} = 103 \\ \sigma _{+} = 50 \\ \sigma _{-} = 50 
\end{cases} \begin{cases}
a = 0.014 \\ n_{4} = 8000 \\ m_{4} = 8000 \\ L_{4} = 137 \\ \sigma _{+} = 50 \\ \sigma _{-} = 50 
\end{cases}  \begin{cases}
a = 0.014 \\ n_{5} = 10000 \\ m_{5} = 10000 \\ L_{5} = 172  \\ \sigma _{+} = 50 \\ \sigma _{-} = 50 
\end{cases}.
\end{split}
\end{equation}
They select the same initial condition of a particle slower than potential wall ($r<\frac{1}{2}$) and we show in Fig.(\ref{fig:polybox}) the evolution of $|\Psi(\alpha,\beta_{\pm})|$ and its full width at half maximum for the first data set.
As in the free particle case, the wave packet spreads with $\alpha$, i.e. it delocalizes until it disappears in a finite $\alpha$ time. The real difference between free particle case and particle in a box case is the trajectory followed by the wave packet. 
If we study the evolution of the wave packet maximum position $\beta_{\pm}^{m}$ for the all data sets, we observe that the wave packet trajectories move away from the \textit{polymer} semiclassical trajectory identified by the initial condition, as we can see in the first graph of Fig.(\ref{fig:picchi}).
The separation from the \textit{polymer} semiclassical trajectory depends on the quantum numbers $n^{*},m^{*}$. In particular, the larger $n^{*},m^{*}$, the longer the semiclassical trajectory is followed. Anyway, no matter how large they are, in a finite time $\alpha$, the wave packetstops following the semiclassical trajectory, is directed to the potential wall and reaches it. As in Fig.(\ref{fig:bounce}), this behavior is repeated for every unexpected bounce against the wall. This way, it is not possible to chose an initial semiclassical state (i.e. large $n^{*},m^{*}$) conserved until the singularity.
This result is opposite respect the one in Eq.(\ref{conservazione}), where in the standard theory the state remains classical until the singularity.
It happens because we have a \textit{time}-dependent initial condition (as in Eq.(\ref{disccon}, it depends on $\alpha$) that changes the particle velocity. 
This behavior is explained if one considers the two different data sets
\begin{equation}
\label{casi2}
\begin{cases}
a_{1} = 0.014 \\ n^{*}_{1} = m^{*}_{1} = 3000 \\ L_{1} = 26 \\ \sigma _{+} = \sigma _{-} = 50
\end{cases} \quad  \begin{cases}
a_{2} = 0.014 \\ n^{*}_{2} = m^{*}_{2} = 400 \\ L_{2} = 26 \\ \sigma _{+} = \sigma _{-} = 50 \end{cases} .
\end{equation}
They respectively select a particle with initial velocity $r<\frac{1}{2}$ and with $r>\frac{1}{2}$. The first one is related to a particle in a box which semiclassically cannot reach the potential wall, while the second one is related to a particle in a box which semiclassically reaches the potential wall. For our purposes, we take two data sets with same values of $a,\sigma_{\pm},L_{0}$ but with different $n^{*}$ and $m^{*}$.
\begin{figure}[h!]
\centering
\includegraphics[scale=.67]{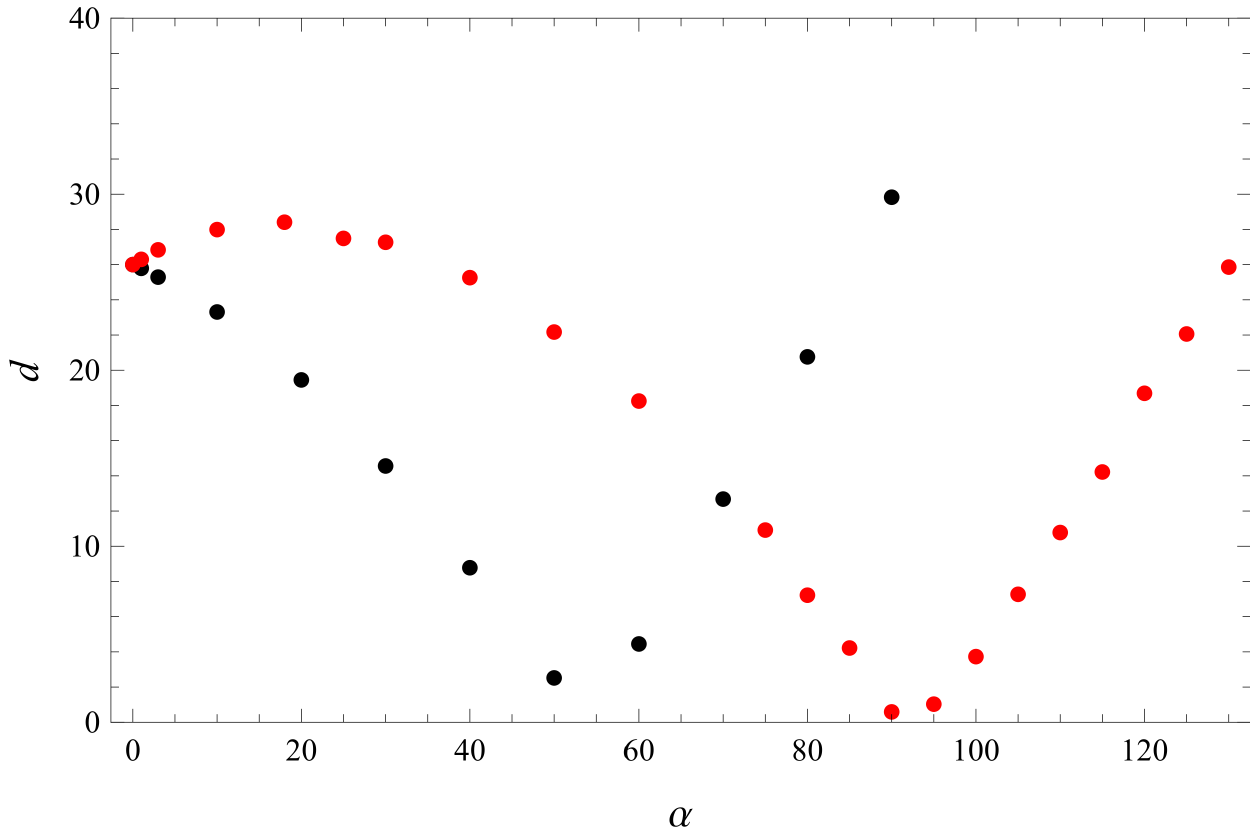}
\caption{\footnotesize "(Color online)". The red(grey) points represent the evolution of the distance $d$ between the wave packet maximum position and the potential wall for $r<\frac{1}{2}$. The black points represent the evolution of the distance $d$ between the wave packet maximum position and the potential wall for $r>\frac{1}{2}$.}
\label{primadopo}
\end{figure}

\begin{figure}[h!]
\centering
\includegraphics[scale=.33]{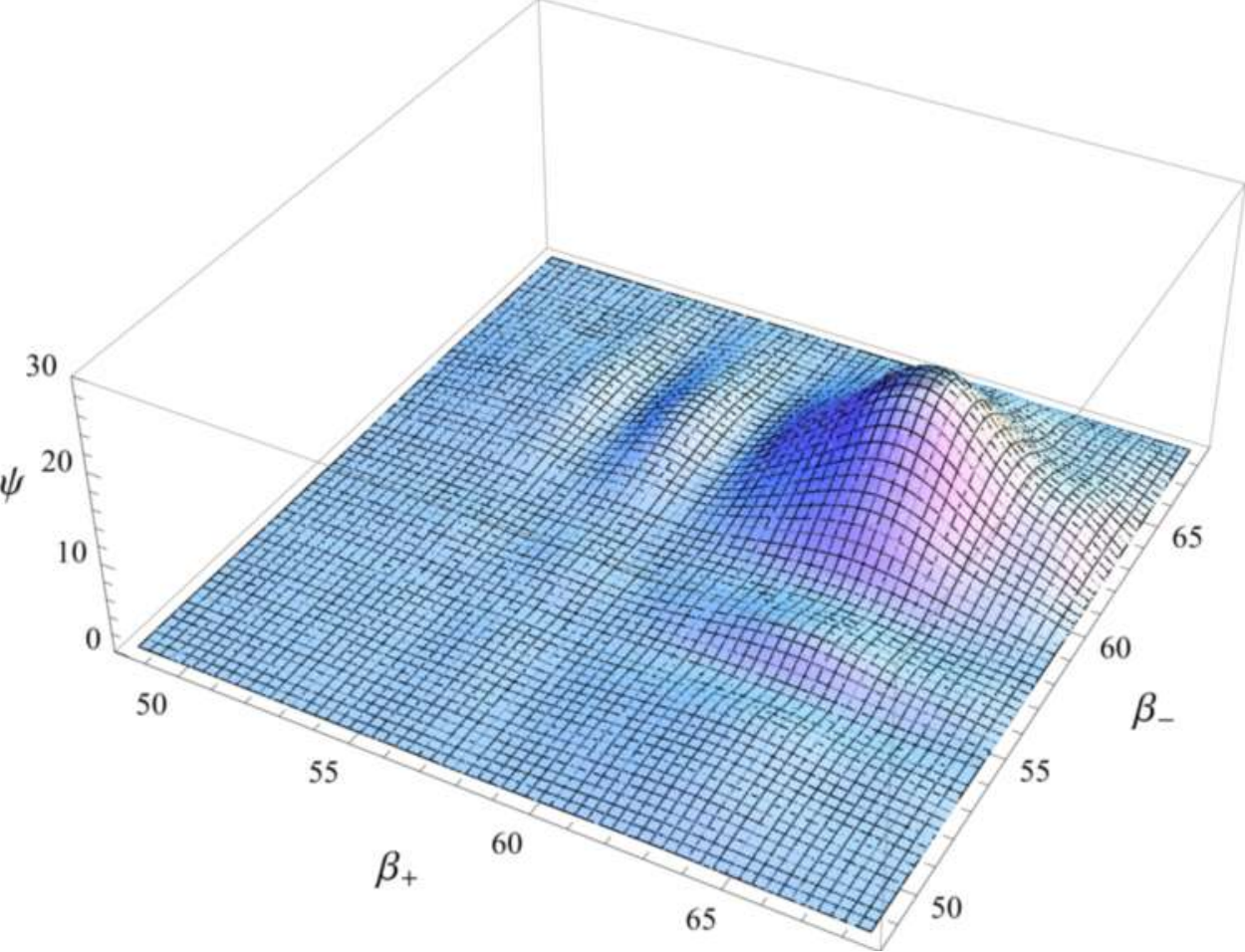}
\includegraphics[scale=.33]{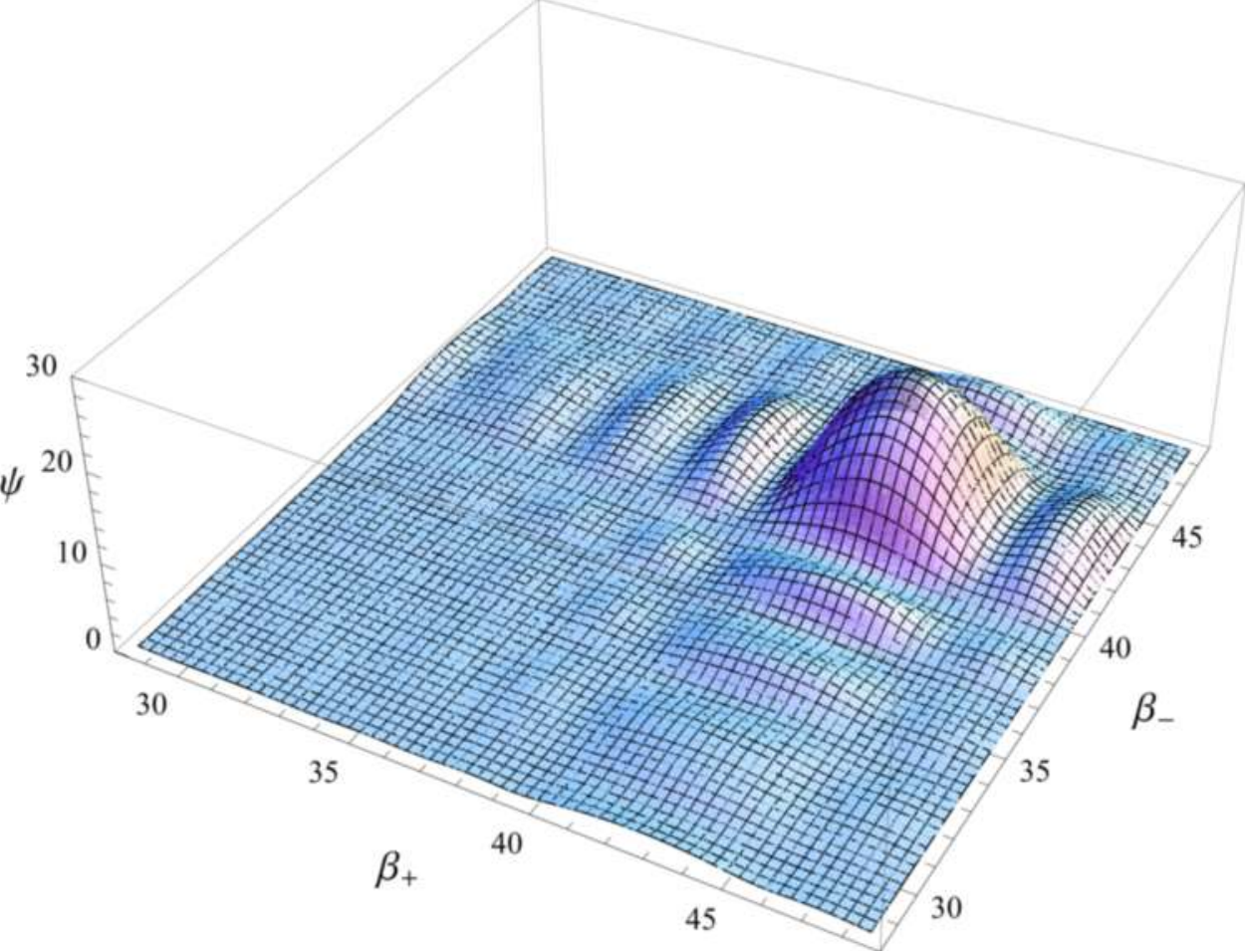}
\includegraphics[scale=.32]{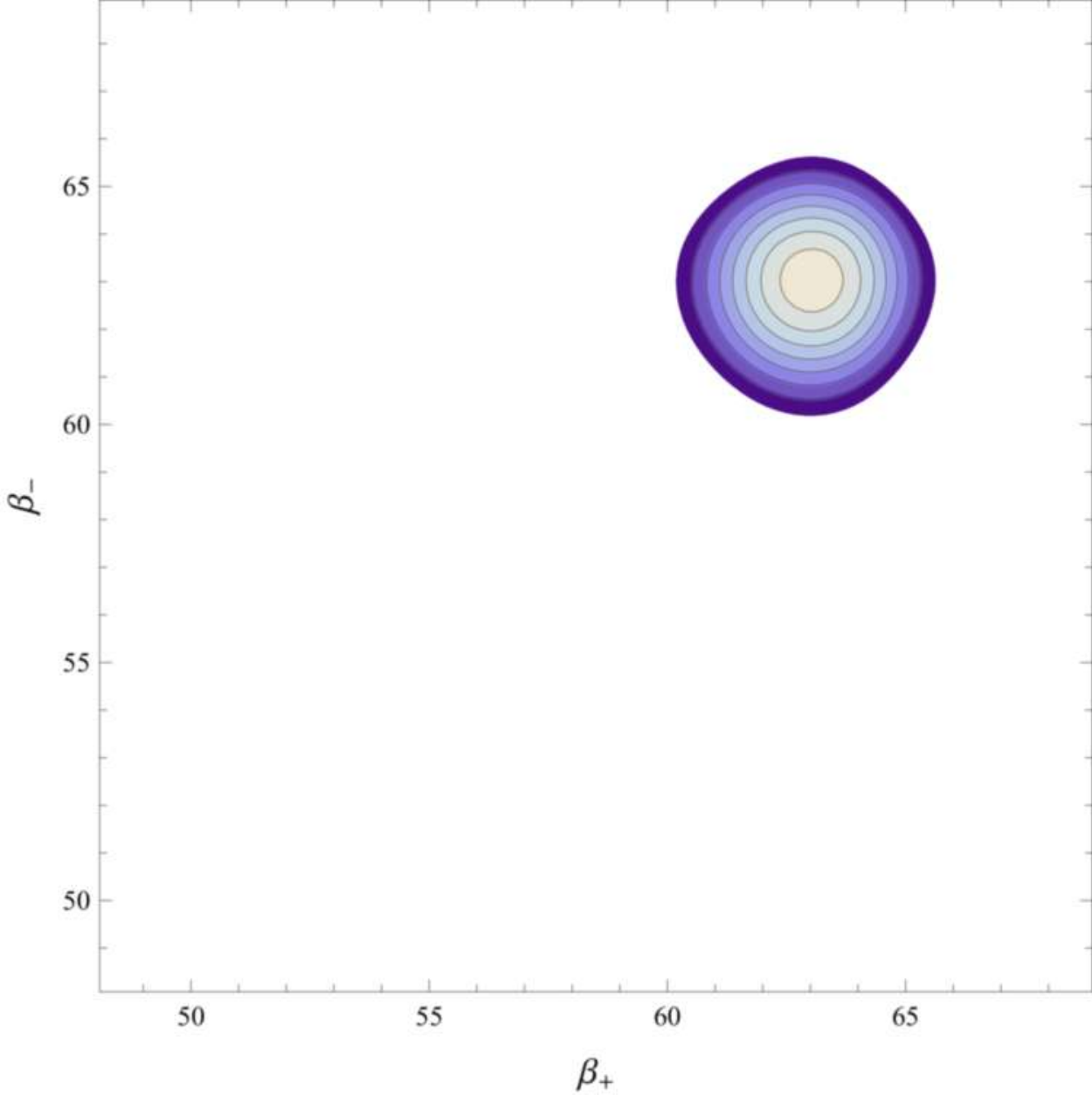}
\includegraphics[scale=.32]{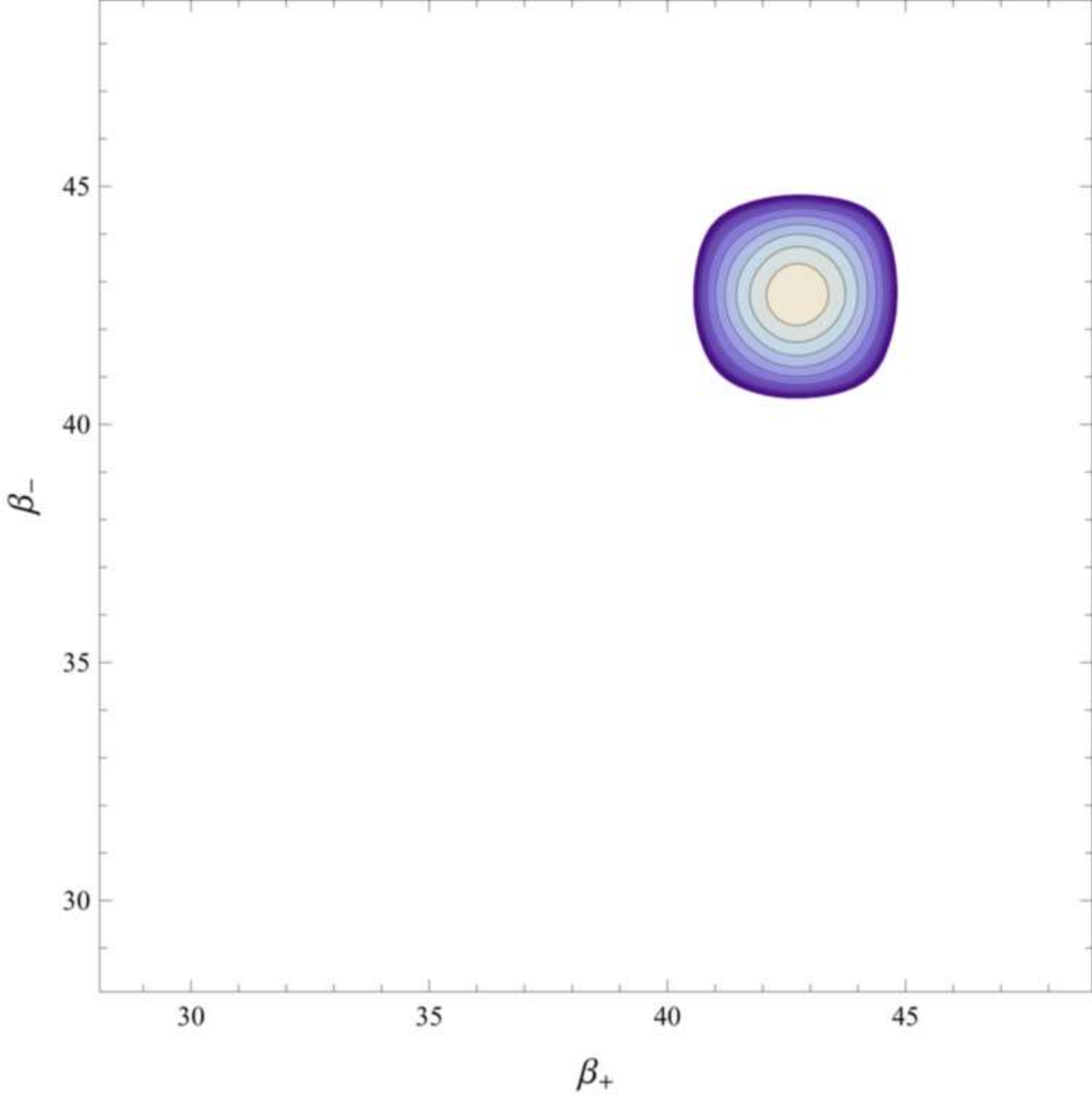}
\caption{\footnotesize "(Color online)". The wave packets and the full width half maximum near the potential wall for the two case with initial condition (\ref{casi2}). The first case is evaluated for $|\alpha|=85$, while the second case is evaluated for $|\alpha|=45$.}
\label{sbatte}
\end{figure}

In Fig.(\ref{primadopo}), the evolution of the distance $d$ between the wave packet maximum position and the potential wall in the two cases towards the singularity is described. \\ When the first one is still traveling, the second one has already bounced on the wall and it is travelling again. The red (light grey) points indicate the (expected) velocity change due to the dynamical initial condition (\ref{disccon}).\\
Finally, it is interesting to study the spread for the two wave pckets near the potential wall. In Fig.(\ref{sbatte}), the two wave packets and the full width at half maximum are sketched. Since the first wave packet should not reach the wall, one would expect a high rate of delocalization near the wall. Instead, as from the second line in Fig.(\ref{sbatte}), the two wave packets near the potential wall have a comparable delocalization. Thus, we can conclude that, when the potential is taken into account as an infinite well, any notion of a free semiclassical wave packet is lost.
\section{CONCLUDING REMARKS}

The Mixmaster model, interpreted as the most general
dynamics allowed by the homogeneity constraint,
constitutes a valuable prototype of the behavior of
a generic inhomogeneous model near the cosmological singularity,
when referred to sufficiently small space regions,
having roughly the causal size.

Therefore, the characterization of its classical
and quantum dynamics has a very relevant value
in understanding the general features of the Universe birth.

The present work is aimed at generalizing \cite{misner},
in which the classical Mixmaster Hamiltonian dynamics
is reduced to the motion of a two-dimensional point-particle
in a closed triangular-like potential and the
corresponding quantum behavior is reconducted to
the one of a point-particle in a box.
The main result of the classical picture is the
neverending bouncing of the particle against the
potential walls (resulting into a chaotic evolution),
while, in the quantum regime, the surprising
feature emerges, of states having very high occupation numbers
which can approach the initial singularity.

This generalization is the
reformulation of the quantum Mixmaster dynamics in the
polymer quantum approach.
We have applied this procedure to the physical degrees
only, i.e. the Universe anisotropies, while the
Universe volume has been kept in its standard interpretation as a time variable for the system evolution.

The semi-classical behavior of the
Mixmaster model, i.e. the classical modified dynamics
by means of the polymer features, results
as chaos-free, in formal analogy with the case
in which a massless scalar field is introduced in
the Einsteinian dynamics. As a consequence, the
quantum regime loses its property to admit very high
occupation numbers asimptotically to the singularity.
Actually, we demonstrated that the absence of a chaotic
behavior prevents to construct the classical
constant of the motion that Misner used to
infer the quantum properties for high occupation numbers. Thus,
the most impressive property of the quantum Mixmaster,
i.e. its ``classicality'' across the Planckian era,
is no longer well-grounded.

In the polymer framework, such impossibility
to recover a quasi-classical behavior near
the singularity, is enforced by noting that it
is impossible to construct wave-packets peaked
around the classical trajectoreies that do not
impact against the potential. Such packets can
follow the classical trajectory for a finite time
interval, after which the bounce of the wave packet against the potential walls takes place.
We showed that this fact is a direct consequence of
the time dependence of the potential well,
resulting in a condition on the free motion of the
wave packets which is correspondingly time dependent and,
soon or later, is violated.

We can conclude that the polymer features of
the Mixmaster model, i.e. the implications of
this particular cut-off physics on the anisotropic degrees
of freedom, enforces the relevance of the quantum
nature of the model near the cosmological singularity,
since they introduce a non-local effect of
the potential walls on the behavior of wave packets,
localized around classical trajectories.
This result suggests that, to better focus on the behavior
of a Mixmaster model model near the cosmological
singularity, it is necessary to implement a more rigurous semiclassical interpretation of the wavefunction toward the cosmological singularity in presence of cut-off induced effects, and a full quantum picture in the disscretized picture.
\begin{acknowledgments}
This work has been partially developed within the framework of the CGW collaboration (http://
www.cgwcollaboration.it).
\end{acknowledgments}

\end{document}